\pdfoutput=1 
\documentclass[fleqn,usenatbib]{mnras}

\usepackage{newtxtext,newtxmath}

\usepackage[T1]{fontenc}

\DeclareRobustCommand{\VAN}[3]{#2}
\let\VANthebibliography\thebibliography
\def\thebibliography{\DeclareRobustCommand{\VAN}[3]{##3}\VANthebibliography}

\usepackage{graphicx}	
\usepackage{amsmath}	
\usepackage{amssymb}	
\usepackage{multirow}

\newcommand{\TESS}{\textit{TESS}}
\newcommand{\ie}{i.e.}
\newcommand{\eg}{e.g.}

\newcommand{\Kepler}{\textit{Kepler}}
\newcommand{\PHOEBE}{\textsc{phoebe\,2}}
\newcommand{\spd}{\textsc{spd}}
\newcommand{\ispec}{\textsc{iSpec}}

\title[DEB in CHTs]{Detached Eclipsing Binaries in Compact Hierarchical Triples: Triple-lined systems BD+44\,2258 and KIC\,06525196}

\author[A. Moharana et al.]
{Ayush Moharana,$^{1}$\thanks{E-mail: ayushm@ncac.torun.pl}
K. G. He{\l}miniak,$^{1}$
F. Marcadon,$^{1,2}$
T. Pawar,$^{1}$
M. Konacki,$^{3}$
N. Ukita,$^{4,5}$
E. Kambe,$^{6}$
\newauthor
and H. Maehara$^{4}$
\\
$^{1}$Nicolaus Copernicus Astronomical Center, Polish Academy of Sciences, ul. Rabia\'{n}ska 8, 87-100 Toru\'{n}, Poland\\
$^{2}$Villanova University, Dept.\ of Astrophysics and Planetary Sciences, 800 East Lancaster Avenue, Villanova, PA 19085, USA\\
$^{3}$Nicolaus Copernicus Astronomical Center, Polish Academy of Sciences, ul. Bartycka 18, 00-716 Warszawa, Poland\\
$^{4}$Okayama Astrophysical Observatory, National Astronomical Observatory of Japan, 3037-5 Honjo, Kamogata, Asakuchi, Okayama 719-0232, Japan\\
$^{5}$The Graduate University for Advanced Studies, 2-21-1 Osawa, Mitaka, Tokyo 181-8588, Japan\\
$^{6}$Subaru Telescope, National Astronomical Observatory of Japan, 650 North Aohoku Place, Hilo, HI 96720, USA\\
}

\date{Accepted XXX. Received YYY; in original form ZZZ}

\pubyear{2022}

\begin{document}
\label{firstpage}
\pagerange{\pageref{firstpage}--\pageref{lastpage}}
\maketitle

\begin{abstract}
Compact Hierarchical Triples (CHT) are systems with the tertiary star orbiting the inner binary in an orbit shorter than 1000 days. CHT with an eclipsing binary as its inner binary can help us extract a multitude of information about all three stars in the system. In this study, we use independent observational techniques to estimate the orbital, stellar, and atmospheric parameters of two triple-lined CHT:  BD+44\,2258 and KIC\,06525196. We find that the masses of stars in  BD+44\,2258 are $1.011\pm0.029 M_{\odot}$, $0.941\pm0.033 M_{\odot}$, and $0.907\pm0.065 M_{\odot}$ while in KIC\,06525196 the estimated masses are $1.0351\pm0.0055 M_{\odot}$, $0.9712\pm0.0039 M_{\odot}$, and $0.777\pm0.012 M_{\odot}$. Using spectral disentangling, we obtained individual spectra of all the stars and combined it with light curve modelling to obtain radii, metallicities and temperatures. Using stellar evolution models from MESA, we constrain the log(age) of BD+44\,2258 to be 9.89 and 9.49 for KIC\,06525196. Two stars in BD+44\,2258 are found to be sub-giants while all three stars in KIC\,06525196 are main-sequence stars. We constrain the mutual inclinations  to certain angles for BD+44\,2258 and KIC\,06525196 using numerical integration. Integrating with tidal interaction schemes and stellar evolution models, we find that KIC\,06525196 is a stable system. But the inner binary of BD+44\,2258 merges within 550 Myrs. The time of this merger is affected by the orientation of the tertiary, even rushing the collapse by $\sim 100$ Myrs when the mutual inclination is close to 90 degrees. 
\end{abstract}




\section{Introduction}

The multiplicity of stars is a well-established phenomenon \citep{multiplicity}. The incidence of multiplicity varies with the spectral type of the stars. Multiplicity is 44\% among the Solar-like stars, out of which $8\%$ are triple stars \citep{solarmultiplesurv}. These numbers increase in O, B, and A types \citep{Mason2009, shatskyntoko, kobulnicky2007}.  There have been a lot of studies towards understanding binaries which has created an almost complete picture of their evolution and formation. The next step in decoding the multiple-architecture is understanding triple systems. 

Binary formation channels can be largely classified into disk instability, core-fragmentation, and N-body interactions. The complexity of formation increases in a triple system with a combination of these formation channels being responsible for their formation \citep{tokoformationreview}. Most of the studies to understand these scenarios using orbital architectures, metallicity variation among stars, and mass distributions in triples, have been usually restricted to wide triples \citep{tokoorbittriples,tokogaiatrip,Leemultipleform}. 

The evolution of triple stars also departs from the simplified evolution of a single star. A triple-star system has the additional complexity of multiple dynamical interactions. The outer companion can directly alter the formation of the components of the inner binary, leading to their apparent difference in properties (\eg apparent age; \citealt{fredv12}), and has been used in explaining various evolutionary phenomena of single and binary stars. 

Hierarchical triple systems have been intensely studied in order to understand the formation of closest main-sequence binary systems \citep{peternlud2001,naozobliqtrip,maxwellkratter}. One of the probable theories of the formation of blue straggler stars involves perturbations from a third body \citep{peretsfab}. One in a thousand high-mass x-ray binaries evolve through interaction with a third star to form low-mass x-ray binaries \citep{eggletonlmxb}. Asymmetry of Planetary Nebulae has been linked to evolution in a triple system \citep{pneakashi,pnebfromtrip} and has even been suspected to play a role in driving white dwarf mergers towards type Ia supernova explosions \citep{maozwdsne}. 

Recent population synthesis studies have shown that 65-75 \% of triples undergo mass transfer \citep{toonen2020}. \citet{masstranslk} have shown that this can occur uniformly throughout the orbit or at certain points due to eccentricity and inclination changes known as von~Zeipel-Lidov-Kozai (ZLK) oscillations \citep{zeipel,lidov,kozai}. Furthermore, these systems perturb Roche-lobe potentials and also undergo Roche-Lobe Over Flow (RLOF), which can occur for the three individual stars and can be circumbinary too. Therefore, understanding the stellar evolution coupled with the dynamical evolution is important when studying triples. 
Most of the known triple systems have long tertiary periods and therefore their dynamical effects can have timescales of decades or centuries. There is a subset of these triples, called Compact Hierarchical Triples (CHT), which offer more potential for observational astrophysics \citep{borkoCHTreview}. The timescale of the changes due to these interactions is short in CHT and can be observed easily over a few years, \eg, vanishing eclipses of HS Hya \citep{hyhya} or re-appearing eclipses of V907 Sco \citep{Zasche_2023}.  These are triples with the outer orbit period shorter than 1000 days. Due to this, dynamical processes in CHTs can be observed in timescales of years. CHTs were thought to be rare \citep{tokostats2004} but with new space-based photometic missions we are discovering more of these systems \citep{rappa2013,borkovitskeptrip}.

Detached Eclipsing Binaries (DEB) are known as the source of the most accurate stellar parameters (\eg, mass, radius, etc.). Accuracy of less than 1\% can be attained by coupling high-precision photometry and high-resolution spectroscopy \citep{torresaccuracy}. The accuracy is robust and independent of different models and methods, even varying slightly due to different numerical implementations \citep{maxted2020,korthmoh}. 

If a CHT has a DEB as its inner binary, there is an added advantage of obtaining accurate stellar parameters of not only the binary but of the tertiary as well \citep{krishides2017}. Using light curve modelling, eclipse timing Variations, spectral analysis, and RVs, we can obtain an accurate picture of the orbits, geometry, stellar parameters, metallicity, age, and evolutionary status. 

There has been a surge in interest in CHT recently. \textit{Kepler} \citep{k2mission} and Transiting Exoplanet Survey Satellite (\textit{TESS}) both have been crucial in detection and analysis of these systems \citep{2015borko,2020borko2tt}. Tertiary stars in CHT have been found to host tidally induced pulsations \citep{fullertidalpulstriples}. Ongoing projects are using triply-eclipsing triples (TET) to characterise CHT \citep{rappa6cht2022}. Studies even show that CHT can produce exotic Thorne-\.{Z}ytkow objects \citep{eisnercht}. Last but not the least, CHT have proven to be useful to study ZLK oscillations and their effect on stellar evolution \citep{borkozlk}. Though most of these studies have provided us with mass and radii of all the stars in a CHT, the tertiary radii-space is dominated by TET or planar systems. 

In this paper,  we report the detection of a DEB in a CHT, BD+44~2258 ($\alpha =$ 13:15:06.66, $\delta = $44:02:33.48; hereafter BD44). BD44 has been previously observed in UV (GALEX; \citealt{galex}) and X-Ray (ROSAT; \citealt{rosat}) but as a single source.  We obtain stellar, orbital, and atmospheric parameters of all the stars in BD44 and a previously detected CHT,  KIC~06525196 ($\alpha =$ 19:30:52.32, $\delta = $41:55:20.81; hereafter KIC65) with \TESS{} photometry and HIDES spectroscopy. We explain the different observations and methods used for extracting parameters in Sections \ref{sec:Observations} and \ref{sec:Analysis}. We use these parameters to estimate the age and evolutionary stages of the components. Further, using the orbital parameters, we  study the evolution and stability of the systems as explained in Section \ref{sec:results}.
\section{Observations}
\label{sec:Observations}
\subsection{Photometry}
We use photometry from \TESS{} \citep{TESSricker} for our light curves\footnote{From GI programmes: G022003, G022062, G04047, G04171, G04234.}(LC). BD44 (TIC 284595199) has 2-minute cadence photometry  obtained from Sectors 16, 22, and 49. KIC65 was a target from the main \Kepler{} mission field, and these data have been analysed in \citet{krishides2017}. In addition to this, the \TESS{} 2-minute cadence photometry (TIC 137757776) is available from Sectors 40 and 41. In this work, we only model a segment of \TESS{} data from Sector 41 for KIC65 while for BD44 we use segments from Sectors 16, 22, and 49. These segments were selected on the basis of spot variability in the LCs. This was also the reason for selecting only Sector 41 for KIC65 because the overall structure of the LC was similar in both sectors but Sector 41 has fewer fluctuations than Sector 40. The structure of the LC for BD44 changes in the sectors and therefore we model all the sectors to check for consistency. 

We filtered out the points which had the best quality-flag for our purposes. There seems to be no other star in the Full Frame Image of BD44 but there seems to be some contamination for KIC65. For BD44, we compared normalised LCs obtained using different apertures on the FFIs only to find they all are identical when normalised. Therefore we consider that any third light that would show up from LC modelling will be solely due to the third star. We considered the Simple Added Photometry or SAP fluxes for modelling both our systems.  

\subsection{Spectroscopy} 
We use spectra collected with the HIDES \citep{izumiurahides,kambe2013}, attached to the 1.88-m telescope at the Okayama Astrophysical Observatory. Observations were conducted in the fiber mode with an image slicer ($R\sim50\,000$), without I$_2$, and with ThAr lamp frames taken every 1-2 hours. The spectra are composed of 62 rows covering 4080--7538~\AA, of which we use 30 (4365--6440~\AA) for radial velocity (RV) calculations. A detailed description of the observing procedure, data reduction and calibrations is presented in \citet{krishides2016}. 

For KIC65 we used exactly the same set of 14 spectra and RV measurements as in \citet{krishides2017}. For BD44 we took a total of 28 spectra.

\section{Analysis}
\label{sec:Analysis}
 In the following sections, we use A-B notation to denote the CHT, where B is the tertiary. A is the eclipsing binary with components Aa and Ab. Aa corresponds to the primary classified according to temperature, usually the deepest eclipse in the LC if not affected by spots. Since we would be talking about six stars in total (three in each of the two CHT), we use the short form for the star's name along with the alphabetical notation to exclusively denote each star, \eg, the secondary (cooler star in the binary) of KIC65 is referred as KIC65Ab.   
\subsection{RV extraction and fitting}
Since both targets were observed with the same spectrograph as part of the same programme, the approach to RV calculations and fitting was essentially identical. It is described in detail in \citet{krishides2017}, but we summarise it here briefly.

The RVs were calculated with a TODCOR method \citep{zuckermazeh} with synthetic spectra computed with ATLAS9 code as templates. Measurement errors were calculated with a bootstrap approach and used for weighting the measurements during the orbital fit, as they are sensitive to the signal-to-noise ratio (SNR) of the spectra and rotational broadening of the lines. Though this code is optimised for double-lined spectroscopic binaries and provides velocities for two stars ($u_1, u_2$), it can still be used in triple-lined systems as well. The RVs of the eclipsing pair were found from the global maximum of the TODCOR map since in both targets these components contribute more to the total flux than the third star. The tertiary's velocities were found from a local maximum, where $u_1$ was set for the tertiary, and $u_2$ for the brighter component of the eclipsing pair. This scheme was used previously for KIC65 in \citet{krishides2017}, and the $rms$ of the resulting tertiary's RVs was comparable to the stability of the instrument estimated from RV standard stars. In some cases, the velocity difference between two components was too small to securely extract their individual RVs, and measurements were taken only for the remaining one. For this reason, each component of a given triple may have a different number of RV data.

The orbital solutions were found using our own procedure called {\sc v2fit} \citep{v2fit}. It applies a Levenberg-Marquardt minimisation scheme to find orbital parameters of a double-Keplerian orbit, which can optionally be perturbed by a number of effects, like a circumbinary body. The fitted parameters are: orbital period $P$, zero-phase $T_P$\footnote{Defined as the moment of passing the pericentre for eccentric 
orbits or quadrature for circular.}, systemic velocity $\gamma$, velocity semi-amplitudes $K_{1,2}$, eccentricity $e$ and periastron longitude $\omega$, although in the final runs, the last two parameters were usually kept fixed to zero. We also included the difference between systemic velocities of two components, $\gamma_2-\gamma_1$, and a circumbinary body on an outer orbit, parameterised analogously by orbital parameters $P_3$, $T_3$, $K_3$, $e_3$, and $\omega_3$. In such case, $\gamma$ is defined in the code as the systemic velocity of the whole triple. 

Systematic errors that come from fixing a certain parameter in the fit are assessed by a Monte-Carlo procedure, and other possible systematics (like coming from poor sampling, small number of measurements, pulsations, etc.) by bootstrap analysis. All the uncertainties of orbital parameters given in this work already include the systematics.

\begin{figure}
    \centering
    \includegraphics[width=0.45\textwidth]{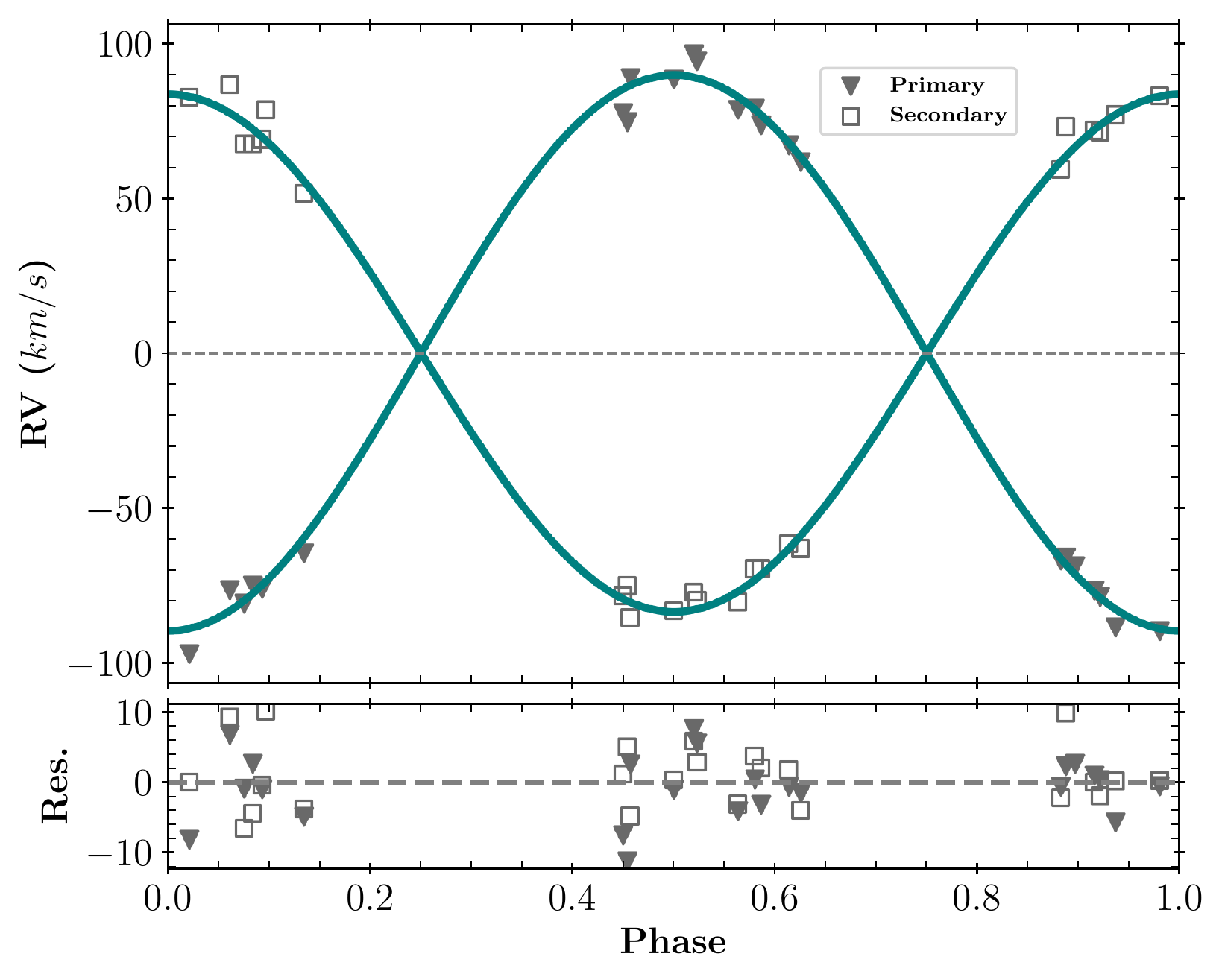}
       \includegraphics[width=0.45\textwidth]{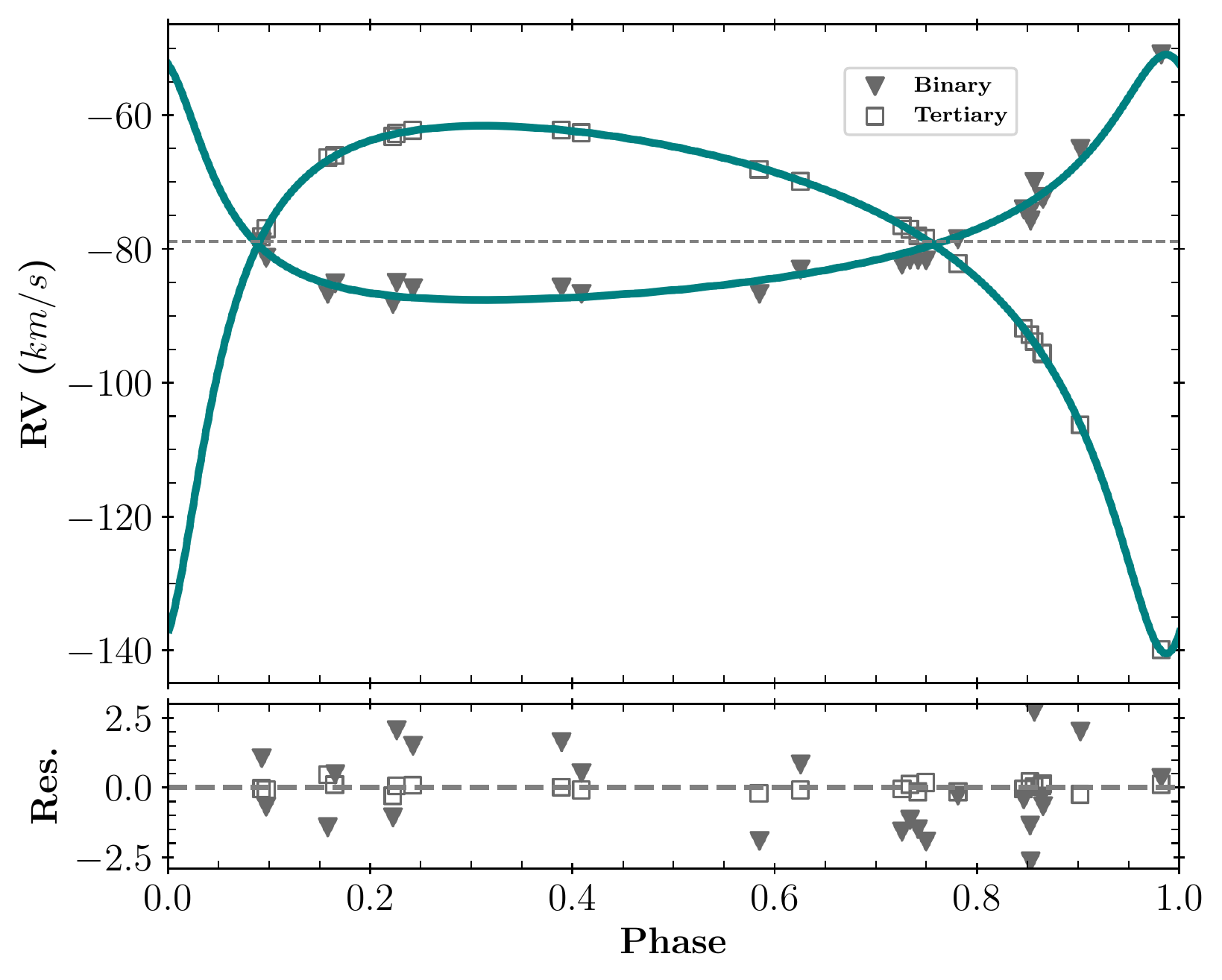}
    \caption{Phased RV profiles of BD44, for the inner binary (top) and the whole system (COM of the binary vs the tertiary). The green lines are the best \textsc{v2fit} models. Though the tertiary has a good fit, the residuals from the binary fit contribute to the final uncertainties in the tertiary parameters.}
    \label{fig:rv_BD}
\end{figure}

In addition, for each observation where three sets of lines were sufficiently separated, we also calculated the systemic velocities $\gamma(t_i)$ of the inner pair, using the formula:
\begin{equation}
\gamma(t_i) = \frac{v_1(t_i)+ q v_2(t_i)}{1+q},
\end{equation}\label{eq_gam}
where $v_{1,2}(t_i)$ are the measured RVs of the inner binary, and $q$ is the mass ratio, found from the RV fit with a circumbinary perturbation. With these values as the center-of-mass (COM) RVs of the binary and RVs of the tertiary component, we can treat the long-period outer orbit as an SB2 (Fig.\ref{fig:rv_BD}), and independently look for its parameters. The final values of $P_3$, $K_3$, $e_3$, etc., actually come from such fits.

\subsection{Broadening Functions}
\begin{figure}
    \centering
    \includegraphics[width=0.45\textwidth]{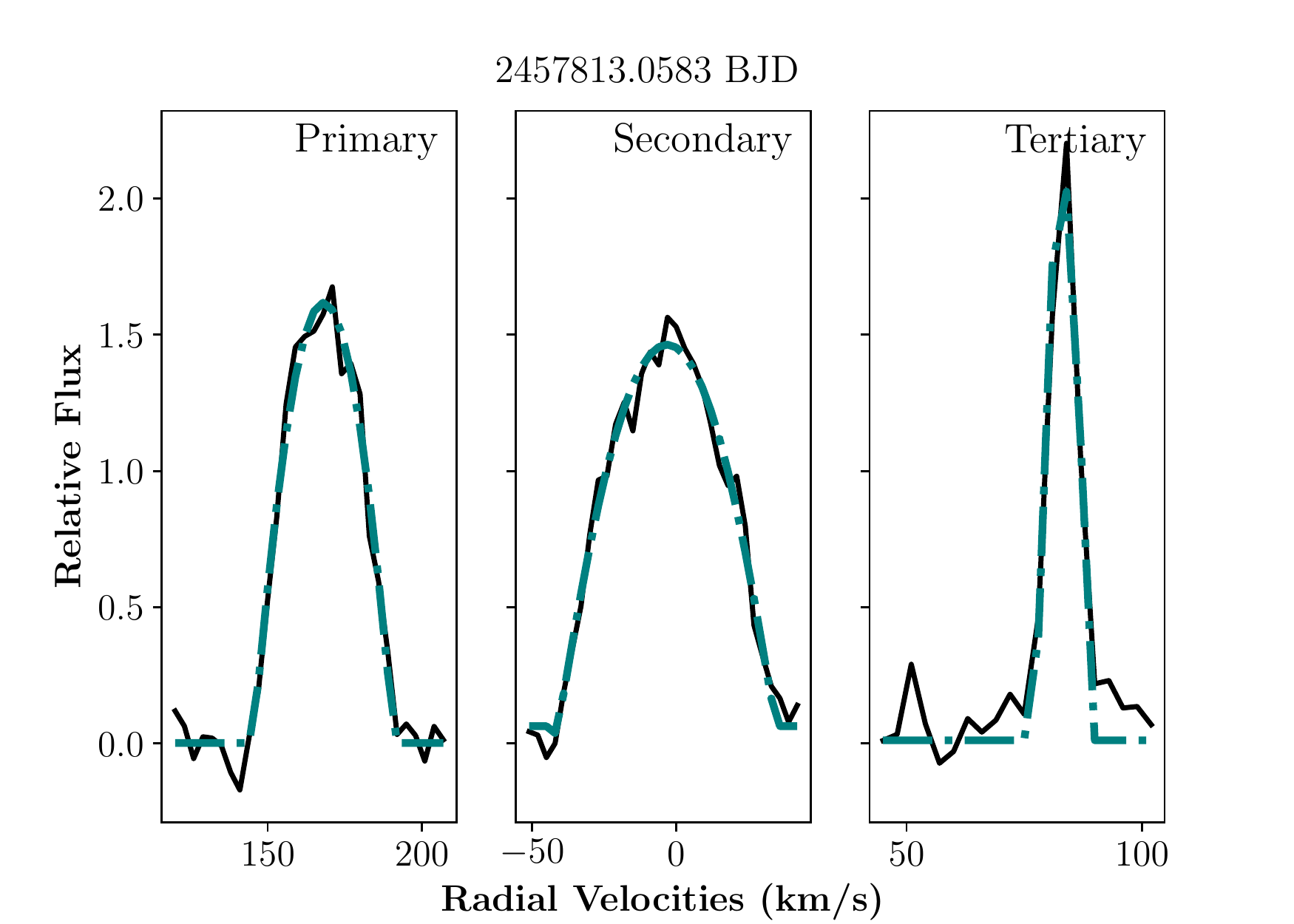}
    \includegraphics[width=0.45\textwidth]{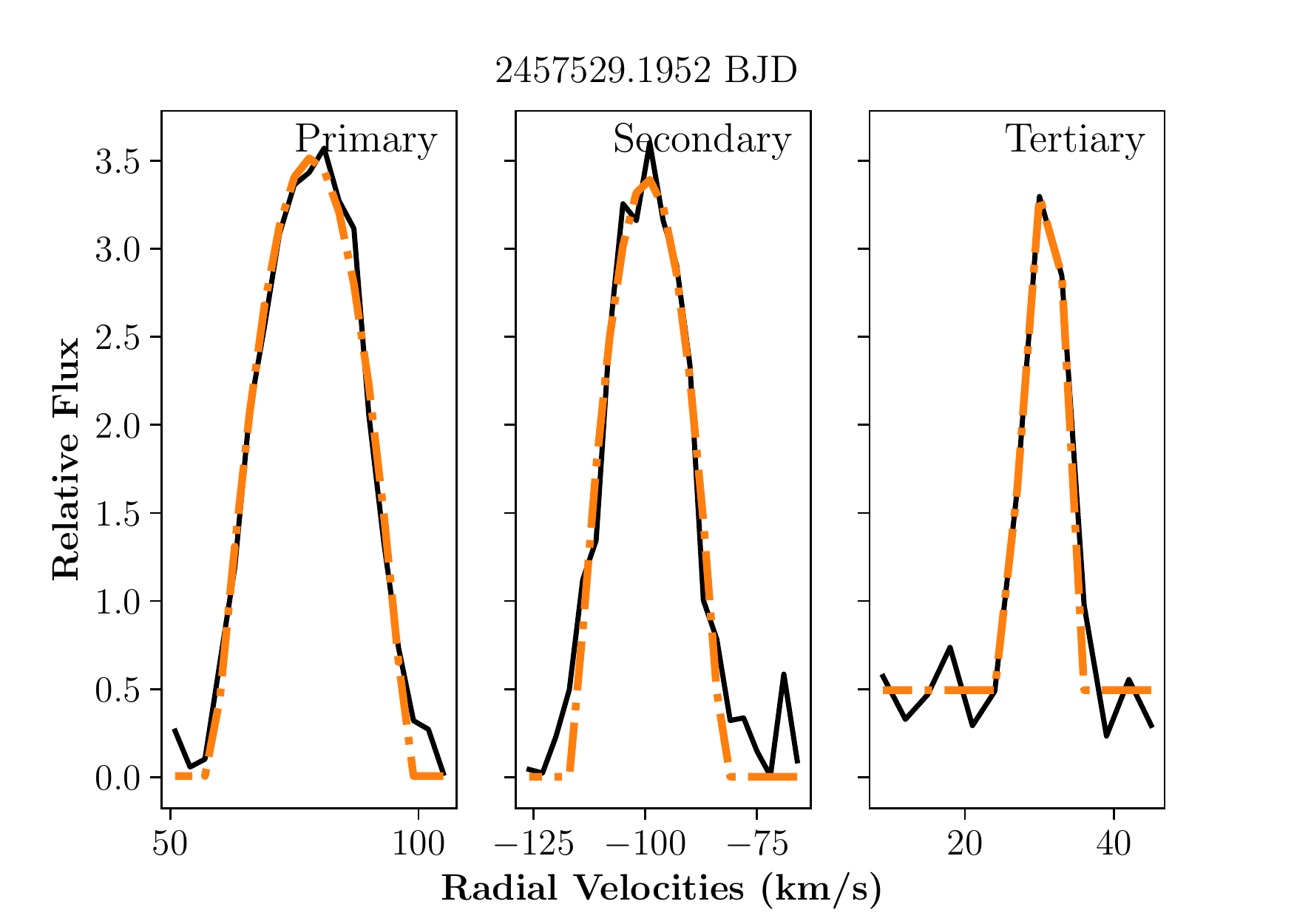}
    \caption{Broadening functions for all the three components of BD44 (top) and KIC65 (bottom). The green and orange dashed line represents best-fit rotational profile for BD44 and KIC65 respectively. The sharp kinks or distortions on the BF peaks are most likely due to spots.}
    \label{fig:BFunction}
\end{figure}

Broadening function (BF) is a representation of the spectral profiles in velocity space. BF contains the signature of the RV shifts of different lines and also intrinsic stellar effects like rotational broadening, spots, pulsations, etc \citep{bf_discussion}.
The BF was calculated using the algorithm described in \citet{bfsvd}. We modified a single-order BF code, \textsc{bf-rvplotter}\footnote
{\url{https://github.com/mrawls/BF-rvplotter}}, to calculate multi-order BF and also fit the function with multiple Gaussian or rotation functions. The BF was calculated in a wavelength range of 5050-5600 \AA. We used a synthetic solar-type spectrum with zero  projected rotational velocity ($v\mathrm{sin}(i)$) as our template. The final BF generated was smoothed with a Gaussian smoother of a 3 $\mathrm{km}\, \mathrm{s^{-1}}$ rolling-window. Three clear peaks were visible in all epochs of spectra which implied that we did not observe the system spectra during eclipses of any of the three stars. The peaks were fitted with the rotational profile from \citet{stellarphotgray},
\begin{equation}
    G(v)=A \left [c_{1}\sqrt{1-\left(\frac{v}{v_\mathrm{max}}\right)^{2}}+c_{2}\left (1-\left (\frac{v}{v_\mathrm{max}}\right)^{2}\right )\right] +l v+k 
\end{equation}
    where $A$ is the area under the profile and $v_\mathrm{max}$ is the maximum velocity shift that occurs at the equator. $c_{1}$ and $c_{2}$ are constants that are a function of limb darkening themselves while $l$ and $k$ are correction factors to the BF ''continuum". The fits revealed distortions (sharp kinks) of the BF from the ideal rotational profiles (Fig.\ref{fig:BFunction}). These are most likely due to spots, and this gives us a qualitative idea about the relative number of spots on the stars. These fits for different epochs were used to (i) calculate light contributions (the parameter $A$) from different companions, and (ii) make an initial estimate of  $v\mathrm{sin}(i)$ (the parameter $v_\mathrm{max}$) for spectral analysis.
We also used BF to calculate RVs and found them consistent with the TODCOR RVs. Therefore, for the sake of consistency and familiarity, we used RVs from TODCOR. 

\subsection{Spectral Disentangling}
\label{sec:specdis}
A detailed study of stellar evolution needs a model-independent  estimate of stellar metallicity. To estimate atmospheric parameters and abundances, we need spectra for all three stars in the CHT.  We use the technique of Spectra Disentangling \citep[hereafter: \spd;][]{simonsttrum_disentangling,hadrava_disent}, for separating individual spectra of the component stars from the composite spectra. This method, though, takes in the assumption that the line profiles are not intrinsically variable. This would mean that we should consider only the out-of-eclipse spectra so as to avoid such variability during the eclipse (\eg, Rossiter-McLaughlin effect). One of the advantages of this method is that it can detect faint companions, \ie, $<3 \%$ light contribution \citep{holmgrenlowlightfrac,2013mayerlowlightfrac}. This, though, requires good phase coverage and high signal-to-noise ratio. 

For our purpose, we used a python-based wrapper\footnote{\url{https://github.com/ayushmoharana/fd3\_initiator}} around the disentangling code \textsc{FDbinary} \citep{fdbinary}, which can disentangle up to three components. The wrapper takes two inputs: (i) an estimate of orbital parameters and (ii) RV corrections and light-ratios at each epoch of spectra used from disentangling. We used the solution from RV fitting as starting values in our optimisation. The light-ratios at every epoch were calculated from the BFs. To make the computation easy and avoid any wavelength dependency, we divided the total spectral range into four sections with overlapping regions. The final disentangled spectra were stitched after normalisation, and removal of the edges of the segmented-disentangled spectra. The overlapping regions acted as check-points for normalisation as they helped us choose the normalisation function which gave the same line-depths for a particular, overlapping spectral line. The errors of the disentangled spectra were taken as the sum of errors calculated from SNR, and flux-scaled residuals from the disentangled routine.

\subsection{Light curve fitting}
We use the version 4 of \PHOEBE{} code\footnote{\url{http://phoebe-project.org/}} \citep{phoebe2016,phoebe2018,phoebeJ2020,conroy2020} for our LC modelling. \PHOEBE{} models eclipsing binaries (or single stars) by discretising the surface of each star. It also distorts the stellar surfaces according to their Roche potentials. The key feature that made us choose this code is its ability to model spots and also solve the inverse problem with spot parameters free for optimisation. 

\begin{figure}
    \centering
    \includegraphics[width=0.45\textwidth]{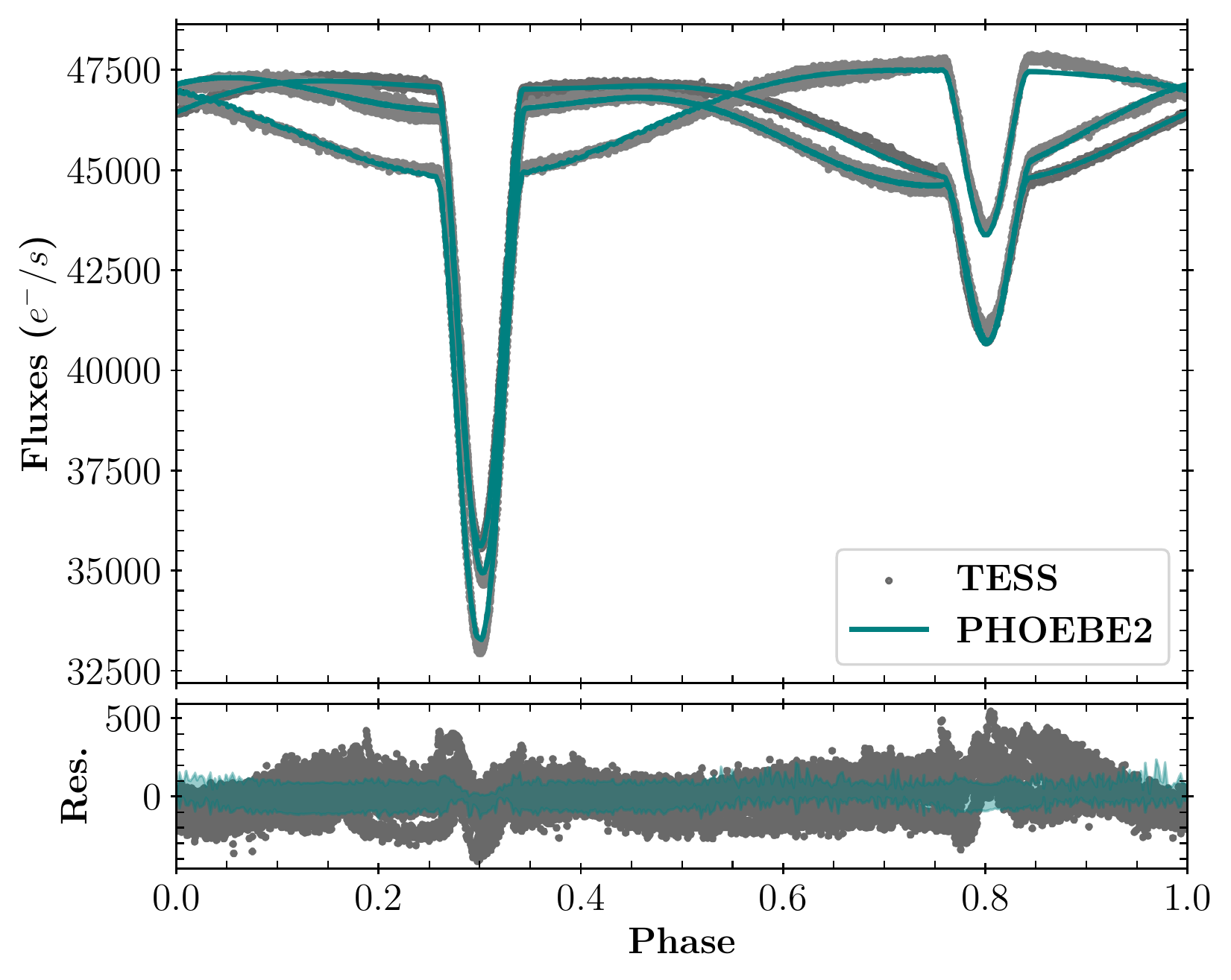}
    \caption{Phased TESS LC (for stable segments from all sectors) for BD44. The green lines represent best-fit \PHOEBE{} model. The shaded areas are translated from the adopted errors of the MCMC solution from Sector-16.}
    \label{fig:ph_lc_BD_all}
\end{figure}

\begin{figure}
    \centering
    \includegraphics[width=0.45\textwidth]{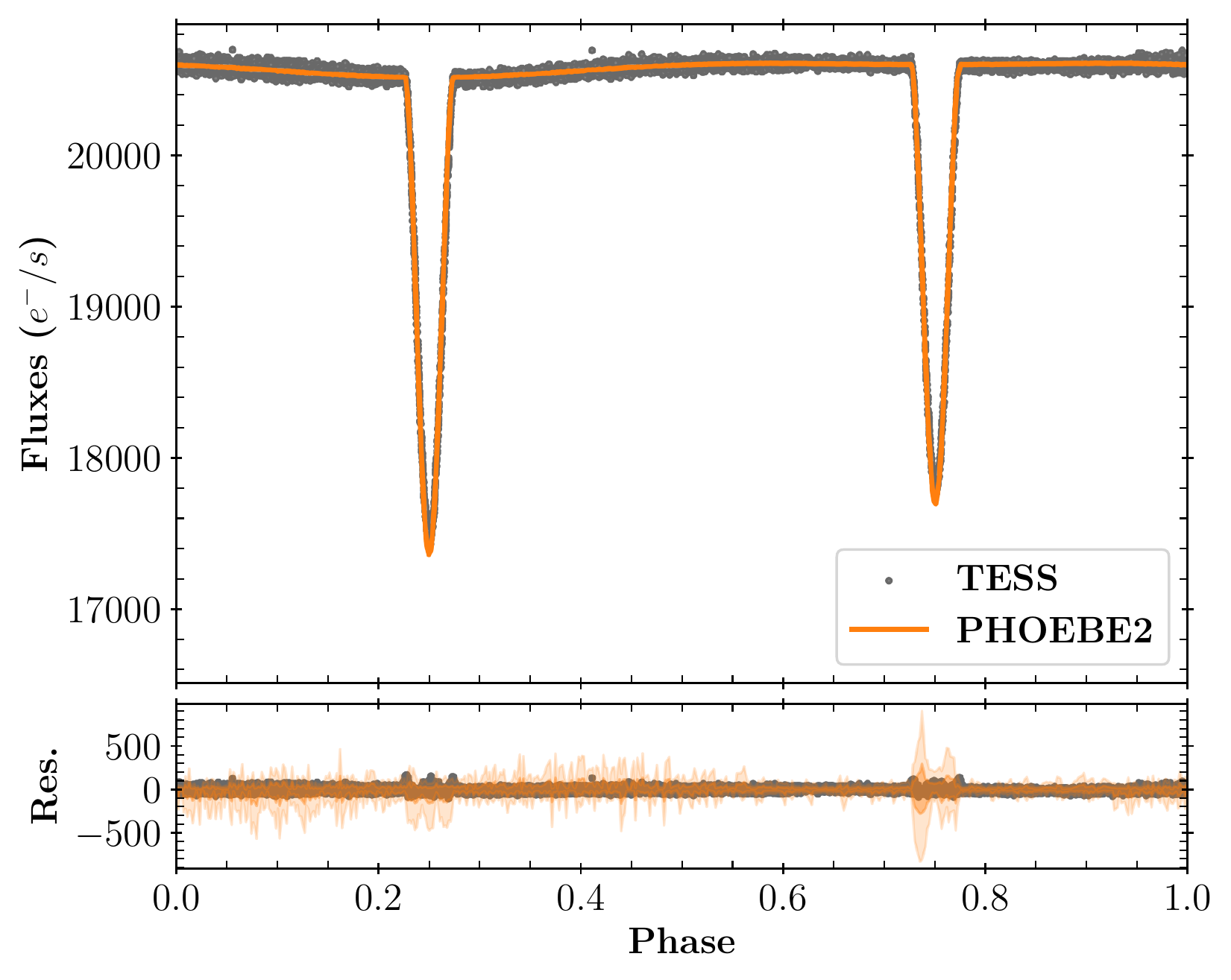}
    \caption{Phased TESS LC for KIC65. The orange lines represent best-fit \PHOEBE{} model. The shaded areas are translated from the adopted errors of the MCMC solution from Sector-41.}
    \label{fig:ph_lc_KIC}
\end{figure}

The first look on the LCs of BD44 and KIC65 suggests the presence of cold spots in the stars. Comparing the light-curve of the different sectors reveals that the spots are time-evolving. We, therefore, approach our modelling by dividing the light-curve into segments with relatively-stable spot signatures. While we model all available sectors for BD44 (Fig.\ref{fig:ph_lc_BD_all}), we only model only Sector-41 for KIC65 (Fig.\ref{fig:ph_lc_KIC}). Further, the distortions on the BF suggest the secondary to be more active than the primary for both stars. Therefore, in our models, we assume more spot(s) on the secondary than the primary. 
The third-light ($l_{3}$) is expected in a triple-lined CHT unless the signatures are removed with detrending methods. The $l_{3}$ is highly degenerate with the inclination ($i$) too. Estimating $l_{3}$ for our systems, from LC, adds another level of complexity with the cold spots affecting the depths of the eclipse.
We assume that the values of light-fractions obtained from BF of optical spectra are similar to the light-fraction of the components in the \TESS{} band. Therefore, we start with an initial $l_{3}$ equal to the flux-fraction of the tertiary from BFs.   

 Considering the above assumptions and fixing parameters obtained from RV fitting, \ie, mass-ratio ($q$), semi-major axis ($a$), and period of the binary($P_{b}$), we used \textit{lc\textunderscore geometry} estimator in \PHOEBE{} for initial parameter estimates. We then added spots one by one and minimising overall trends in the residuals manually. We added 1 spot on the secondary of KIC65 and stopped at three spots (2 on secondary, 1 on primary) for BD44.  More spots would have compromised our computational resources. We then optimised for stellar parameters using Nelder-Mead \citep{neldermead} optimisation module in \PHOEBE. The optimised parameters include radii of primary and secondary ($R_\mathrm{Aa}$ and $R_\mathrm{Ab}$ respectively), time of super-conjugation ($T_{0}$), ratio of secondary temperature to the primary temperature ($T_\mathrm{ratio}$), $P_{A}$, $l_{3}$, $i_{A}$, and passband luminosity of the primary ($L_\mathrm{Aa}$). Since the RV fitting didn't show any substantial eccentricity for the binary ($e_{A}$), we kept it fixed at zero which also reduced the computational cost.  After minimising the residuals below 1\% of the total flux, we optimised for all spot parameters (co-latitude\footnote{Co-latitude is measured along the spin axis with the North pole as $0^{\circ}$.} $c^\mathrm{spot}$, longitude $l^\mathrm{spot}$, relative-temperature $T^\mathrm{spot}$, and radius $r^\mathrm{spot}$) for all the three spots. We then randomly optimised, combinations of spot parameters and stellar parameters, to check the robustness of the optimisation.
 
 We calculated the errors through the Monte Carlo Markov Chain (MCMC) sampling, implemented in \textsc{emcee} \citep{emcee2013,emcee2019} and available as a sampler in \PHOEBE. We decided on 8 parameters to be sampled including the relative temperature of the biggest spot in both the systems ($T^\mathrm{spot}_{Ab}$). Due to the lack of computational resources, our initial sampling consisted of 40 walkers and was sampled till we got stable chains for 1000 iterations. Then we started from this sample-space and ran another sampling for 3500 stable chains with 80 walkers. To further check the convergence, we use auto-correlation plots \citep{autocorrtests} with Bartlett's formula \citep{bartlett} to find the minimum limit for auto-correlated chains. The uncertainties represent 68.27\% confidence interval. To validate our results, we translated the errors on the light curves and also the residuals (Fig. \ref{fig:ph_lc_BD_all} and \ref{fig:ph_lc_KIC}). 

\begin{figure*}
   \centering
    \includegraphics[width=\textwidth]{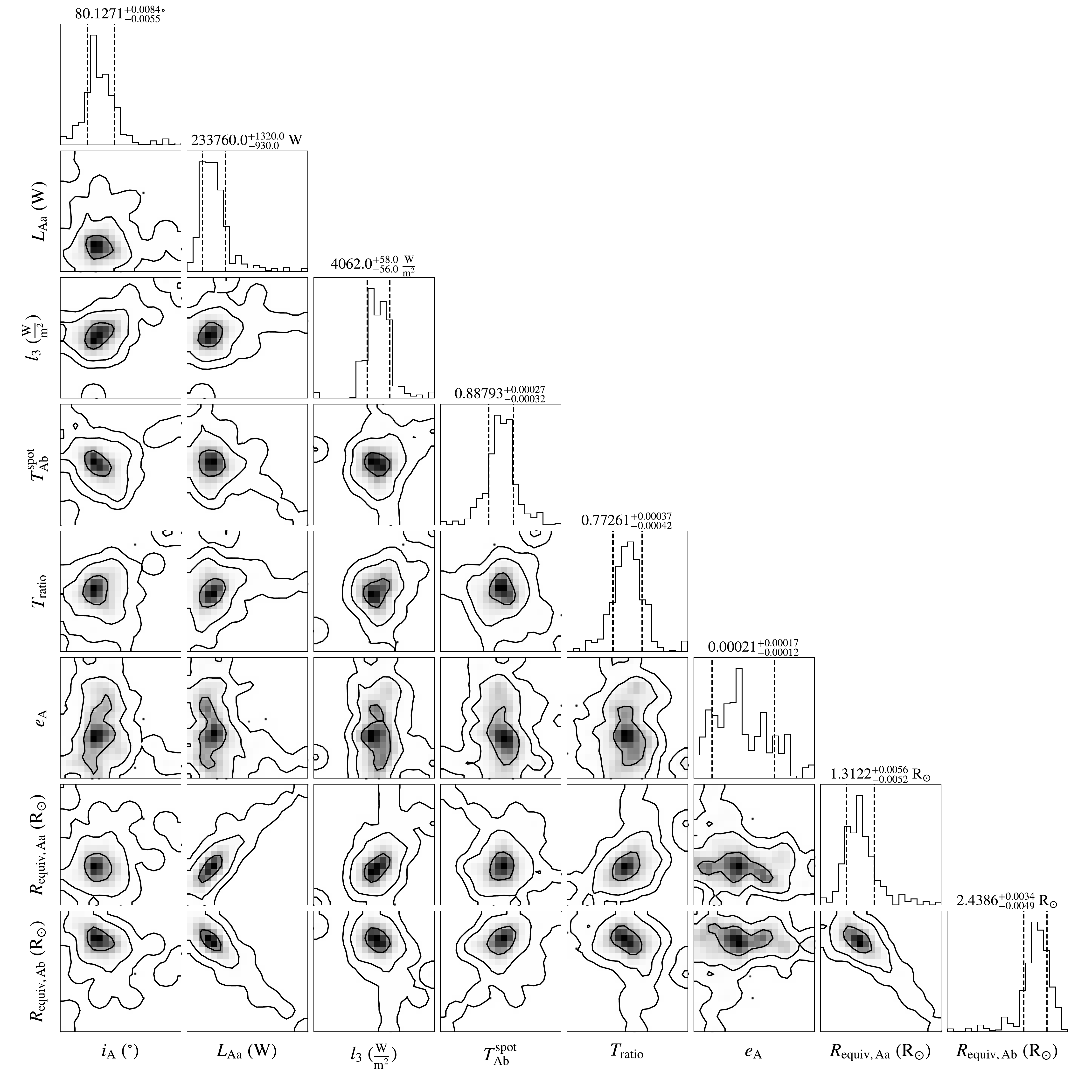}
   \caption{Corner plot for MCMC parameters from LC fitting of BD44. The contours on the maps correspond to 1$\sigma$, 2$\sigma$, and 3$\sigma$ errors on the distribution. The effect of spots is visible in the distorted maps of the eccentricity parameter-space.}
   \label{fig:ph_mcmc_BD}
\end{figure*}

\begin{figure*}
    \centering
    \includegraphics[width=\textwidth]{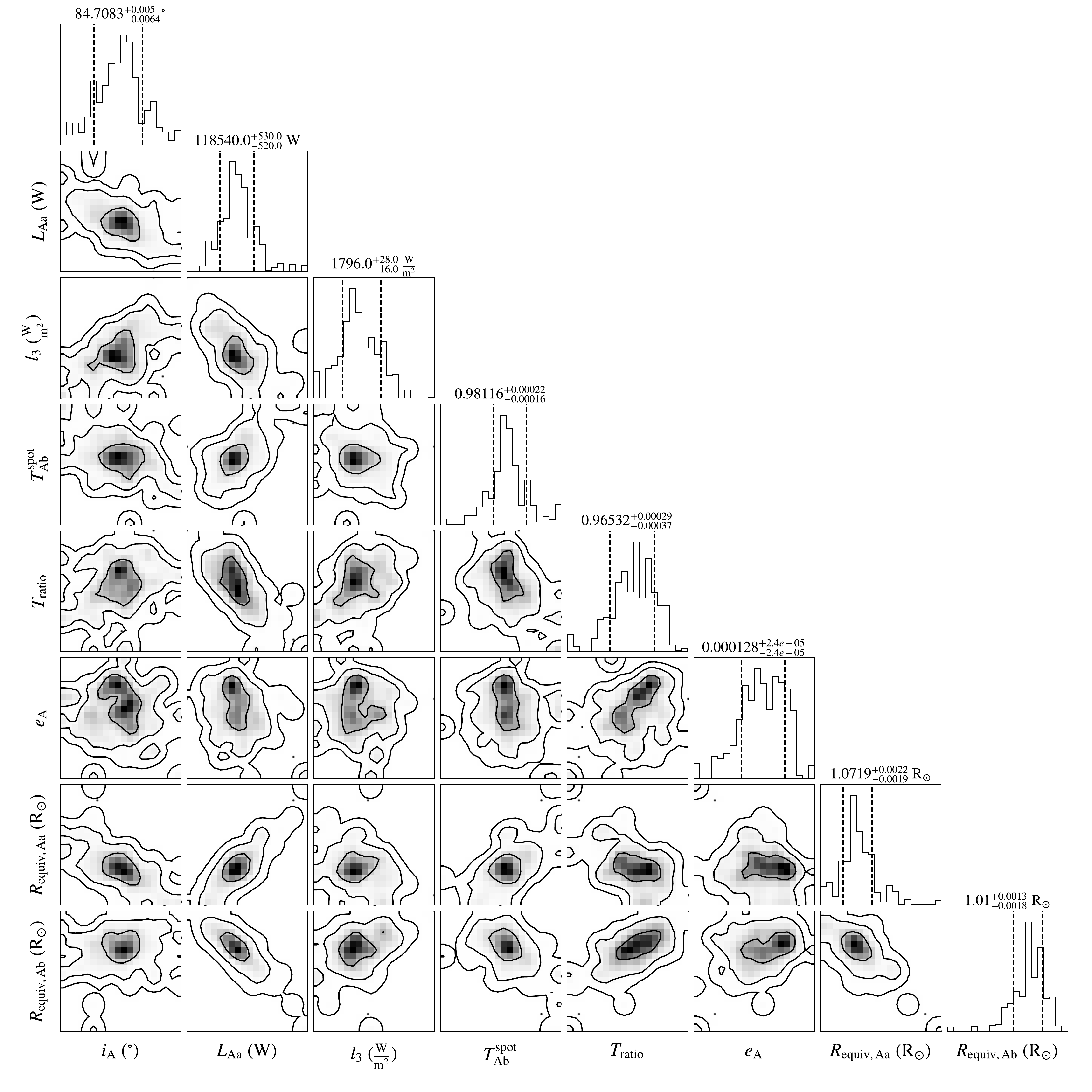}
     \caption{Corner plot for MCMC parameters from LC fitting of KIC65. The contours on the maps correspond to 1$\sigma$, 2$\sigma$, and 3$\sigma$ errors on the distribution. The effect of spots is visible in the distorted maps of the eccentricity parameter-space. Also note that the distortion of the radii parameter-space is smaller than that of BD44.}
    \label{fig:kicmcmc}
\end{figure*}

\subsection{Spectroscopic Analysis}
\begin{figure*}
    \centering
    \includegraphics[width=\textwidth]{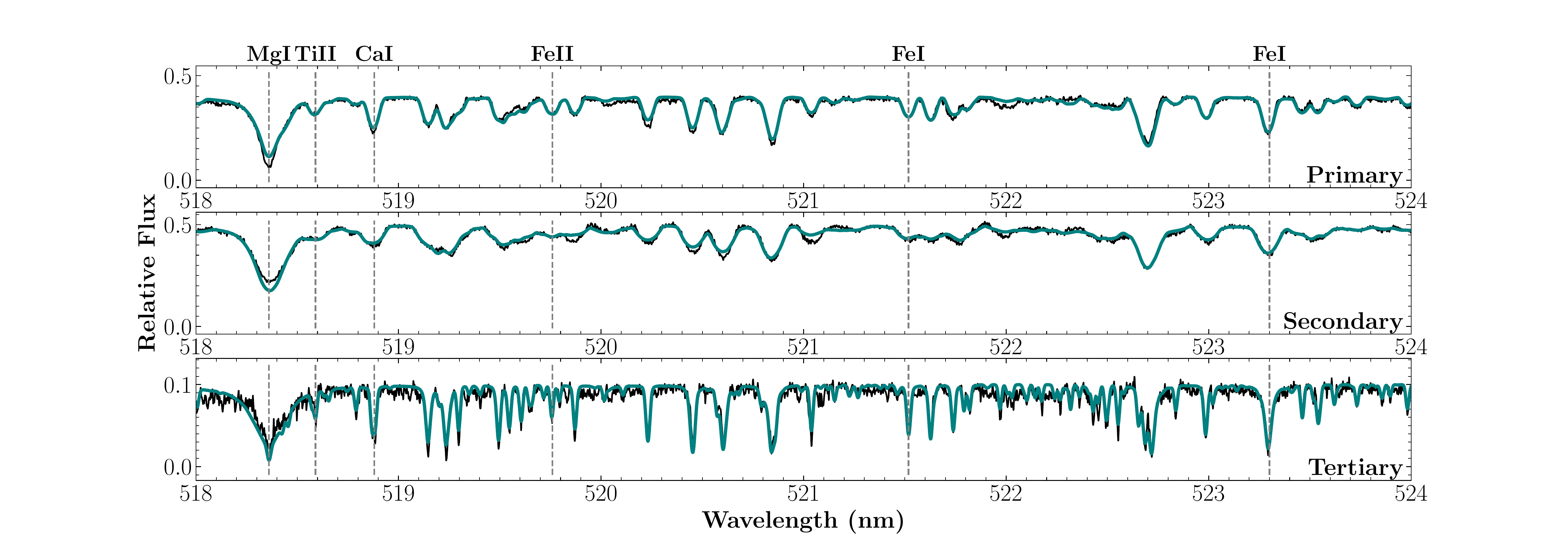}
        \includegraphics[width=\textwidth]{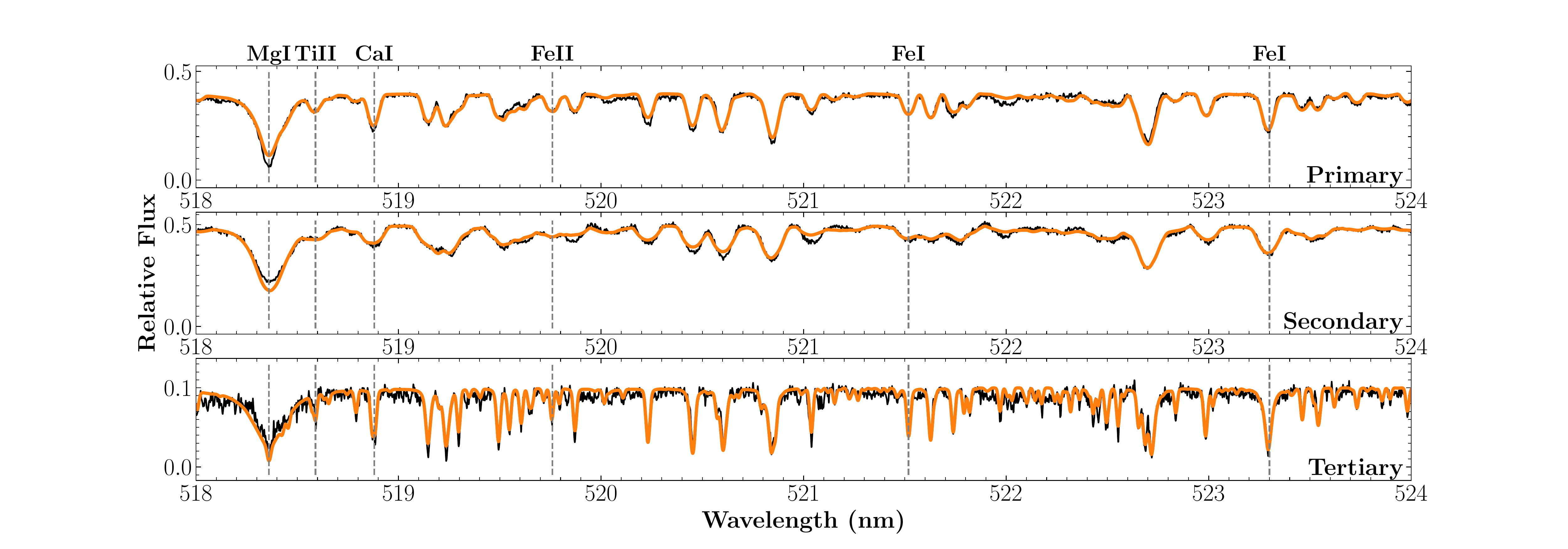}
    \caption{A section of the disentangled spectra and the best-fit synthetic template of all the components of BD44 (top) and KIC65 (bottom). The fit was obtained through the SSF method available in \ispec. While the SNR of the primary and secondary vary between 50-60, the tertiary component has a SNR $\sim$18. The section contains one of the wings of the Mg triplet and several Fe lines.}
    \label{fig:spectrafitsBD}
\end{figure*}

We used \ispec{} \citep{ispec2014,ispec2019} for spectroscopic analysis. Before the analysis, we prepared the spectra by correcting the RV offsets that was present in the \spd{} products. We fitted and corrected the continuum for the \spd{} spectra using a spline function of degree 3, consisting of 30-150 splines, depending on the component. We used the synthetic spectral fitting (SSF) method where \ispec{} generates synthetic spectra on the go for the fitting. We used line masks for specific regions, generated by the spectral synthesis code \textsc{spectrum}\footnote{\url{http://www.appstate.edu/~grayro/spectrum/spectrum.html}} to calculate the $\chi^{2}$. The regions were generated on the line lists from Gaia-ESO Survey (GES; \citealt{gso2012,gso2013}) version 5.0 which covers the wavelength range of 420nm-920nm. There are two sets of GES line lists available in \ispec, which we use: (i) list for abundances estimates ($\mathrm{LL_{a}}$) and (ii) list for parameter estimate ($\mathrm{LL_{p}}$).   The spectral synthesis was again done using \textsc{spectrum} with model atmospheres from \citet{marcges}. The solar abundances were chosen from \citet{grevesse2007}. In our initial test runs, with all parameters free, we found that the $v\sin{i}$ values were consistent with the values obtained from BFs. Therefore in our further runs, we fixed the respective $v\sin{i}$ for the components. Using $\mathrm{LL_{a}}$, we did a search for metallicity, [M/H], on all three stars of the two CHT. Since the [M/H] of the stars of the respective CHT were consistent within the uncertainty range, we averaged the [M/H] estimates for the system and fixed it for further runs. In the final runs, we fixed the $\log(g)$ (obtained from LC models) for primary and secondary. The limb darkening coefficient (values were taken from \citealt{claretlimbdarkening}) and resolution were kept fixed  while  marco-turbulence velocity was calculated automatically from an empirical relation established by GES and built in the code. For the estimation of temperatures (and $\log(g)$ for the tertiary), we selected only the part of the spectra with SNR of more than 40 (more than 18 for the tertiary). We chose $\mathrm{LL_{p}}$ for this fitting. The best fit synthetic spectra for all the stars in the study are shown in Fig.\ref{fig:spectrafitsBD}.  We also calculated the radii $R$ (in R$_\odot$) of the tertiary stars using $\log(g)$ (in dex) from spectra, by applying the formula:

\begin{equation}
    R = A_{c}\sqrt{\frac{M}{10^{\log (g)}}},
\end{equation}

where $M$ is the mass of the tertiary calculated from LC and RV fitting (in M$_\odot$) and $A_c \equiv \sqrt{G M_\odot}/R_\odot (= 168.589888477)$ is a constant necessary for transformation to solar units.

We calculated the $\alpha$-enhancement with $\mathrm{LL_{a}}$ and fixed the rest of the parameters as given in Table.\ref{tab:paramtable}. Using this setup, the abundances\footnote{The abundances were obtained in the 12-scale as $A(X)$, where, $A(X)=\log\left(\frac{n_X}{n_H}\right)+12$, in which $n_X$ and $n_H$ are the number of atoms of the element X and of hydrogen, respectively.} were calculated using the SSF method but with free abundance and [M/H] for a particular element. We did not consider the abundances of the elements where we got large errors and/or where [M/H] was out of the pre-calculated bounds.  

\subsection{Isochrone Fitting}

The age of each system was estimated with a grid of isochrones generated using a dedicated web interface,\footnote{\url{http://waps.cfa.harvard.edu/MIST/}} based on the Modules for Experiments in Stellar Astrophysics \citep[MESA;][]{mesa2011,mesa2013,mesa2015,mesa2018}, and developed as part of the MESA Isochrones and Stellar Tracks project \citep[MIST v1.2;][]{mistc2016,mistd2016}. The grid was prepared for iron abundance, [Fe/H]\footnote{It is reasonable to assume that without significant deviations from solar amounts of $\alpha$-elements, the iron abundance [Fe/H] sufficiently approximates the amount of metals [M/H].}, values from -4.0 to 0.50~dex with 0.05~dex steps, as well as for ages 10$^{8.6}$ to 10$^{10.2}$~Gyr, in logarithmic scale, every $\log(\mathrm{age})$=0.01.

On each isochrone we were looking for a triplet of points that simultaneously best reproduce the observed values of any selected parameters from the following: masses $M_{1,2,3}$, radii $R_{1,2,3}$, and effective temperatures $T^{\rm eff}_{\rm 1,2,3}$ of three components, flux ratio of the inner binary's components $l_2/l_1$ in the given band, as well as the [Fe/H], and distance $d$. The method of simultaneously obtaining distances and reddening $E(B-V)$ reproduced by a given model (triplet of points on an isochrone) is described in Appendix~\ref{appx:isochrones}, as well as in \citet{hel21}. In the recent 3rd {\it Gaia} Data Release \citep[GDR3;][]{gaia2022}, solutions from the {\it Part~1. Main Source} have high value of the RUWE parameter ($\sim$2.6), and better solutions, with significantly different distances, are presented in {\it Part~3. Non-single stars}. We used the latter ones as the constraints in our isochrone fitting. 

It is worth noting that the {\sc ispec} value of [M/H] was also used as a constraint, and the ``best-fitting'' [Fe/H] was searched for in the fitting process. For this reason, the values of [Fe/H] (assumed equivalent to [M/H], since [$\alpha$/Fe]=0) may not be the same as [M/H] found from spectra. In the literature, isochrone or evolutionary track fitting is often made under the assumption of fixed [M/H]. We find this approach incorrect.

\subsection{Numerical Integration of Orbital Dynamics}
\label{sec:numitegration}
The accurate orbital parameters obtained enable us to probe the significant dynamical changes in these compact systems. The major orbital parameters that drive the dynamics are usually the masses, semi-major axes or periods, and all inclinations, including mutual inclination ($i_m$). While we get almost all orbital parameters from combined LC and RV analysis, we still lack the information about the longitude of ascending nodes ($\Omega_{\mathrm{A}}$ and $\Omega_\mathrm{AB}$ for inner and outer orbit respectively) and $i_m$. An estimate of the range of the $i_m$  can be calculated from the constraints arising due to geometry, using $i_\mathrm{A}$ and $i_\mathrm{AB}$. Simplifying the calculations from \cite{mut_i_from_i1_i2} we get,

\begin{equation}
    \cos{i_m} = \cos{(\Omega_\mathrm{A}-\Omega_\mathrm{AB})} \times \sin{i_\mathrm{A}}\sin{i_\mathrm{AB}} + \cos{i_\mathrm{A}}\cos{i_\mathrm{AB}}
\end{equation}
Since the value of $\cos{(\Omega_\mathrm{A}-\Omega_\mathrm{AB})}$ can vary between -1 and 1, we estimate the range of $\cos{i_m}$ to be,
\begin{align}
        \cos{i_m} \geq \cos{i_\mathrm{A}}\cos{i_\mathrm{AB}}-\sin{i_\mathrm{A}}\sin{i_\mathrm{AB}}  \\
        \cos{i_m} \leq \cos{i_\mathrm{A}}\cos{i_\mathrm{AB}}+\sin{i_\mathrm{A}}\sin{i_\mathrm{AB}}
\end{align}
Using trigonometric identities, we get the range of $i_m$ to be,
\begin{equation}
    i_\mathrm{A} - i_\mathrm{AB} \leq i_m \leq i_\mathrm{A} + i_\mathrm{AB}
\end{equation}

This is the solution when the cosine of $i_m$ is positive (say configuration A). There exists another set of solution for a negative cosine configuration (say configuration B),
\begin{equation}
    180-(i_\mathrm{A} + i_\mathrm{AB}) \leq i_m \leq 180 - (i_\mathrm{A} - i_\mathrm{AB})
\end{equation}
This gives us a range of $0^{\circ}-169.4^{\circ}$ (for Config.A) or $10.6^{\circ}-180^{\circ}$ (Config.B) for the $i_m$ of KIC65. While for BD44, we get a range of $3.98^{\circ}-156.27^{\circ}$ (Config.A) or $23.73^{\circ}-176.02^{\circ}$ (Config.B).
To further constrain this range, we rule out the unrealistic $i_m$ by looking at average $i_\mathrm{A}$ variations in the numerical integration of orbital parameters, and comparing with the $i_\mathrm{A}$ from the observations.  For our work, we use \textsc{rebound}\footnote{\url{https://github.com/hannorein/rebound}}, an open-source collisional N-body code \citep{rebound}. \textsc{rebound} can also be used to simulate collision-less problems such as the three-body hierarchical orbit in our case. We use the symplectic integrator \textsc{whfast} which is designed for long-term integration of gravitational orbits \citep{whfast}. \textsc{whfast} uses mixed-variables (Jacobi and Cartesian) and also a symplectic corrector which ensures accurate and fast integration of the N-body dynamical equations. The general setup is defined by masses and orbital parameters obtained with our observations. The values of inclinations, $\omega$ and other orientation parameters are also taken from our observations (Table.\ref{tab:paramtable}).  Further, we use the \textsc{reboundx} \citep{reboundx} library for adding in tidal forces and dissipation to this setup. The implementation uses a constant time-lag model from \citep{hut1981} to raise tides on the larger (and more massive) stars. The constant time-lag parameter ($\tau$) is given by,
\begin{equation}
\centering
    \tau =\frac{2R^{3}}{GMt_{f}}
\end{equation}
 where $R$, $M$ are the radius and mass of the body with tides (BD44Ab and KIC65Aa in our case) and $t_f$ is  convective friction time which is assumed as 1yr as described in  \textsc{reboundx} setup\footnote{We followed the process of adding tides as explained in: \url{https://github.com/dtamayo/reboundx/blob/master/ipython_examples/TidesConstantTimeLag.ipynb}}. $G$ is the gravitational constant. The other inputs from observations include the rotation frequency of the secondary and its radius. The tidal Love number of degree 2 (\texttt{tctlk2}) is assumed to be 0.01. Using this setup, we simulate the systems for a time equal to the time between the observation of the first and last LC of the corresponding systems, \ie, 2.7yrs for BD44 and 6.9yrs for KIC65. We track the inclination changes of the inner binary for different values of $\Omega_\mathrm{AB}$ (and assuming $\Omega_\mathrm{A}=0^{\circ}$) and then compare it with the observations. \\
 A long-term look at such close systems ($P_{A}\sim 3 d$) would require a coupled stellar evolution and dynamical evolution treatment. For this, we use the \texttt{parameter-interpolation} module in \textsc{reboundx}. We update the radius and masses in the numerical simulation from MIST evolutionary tracks for the tidal stars (BD44Ab and KIC65Aa) in the inner binary. We start with an integration timestep of 1/30 times the inner binary period but update the integration timestep, if the orbit shrinks, so that we can track close encounters or collisions. With this setup, we simulate our systems for 600 Myrs.

\section{Results and Discussion}
\label{sec:results}
\subsection{Physical parameters of the Binary}
The binary components of the two systems are found to be of similar orbital configurations. This gives us an interesting case of comparing the effect of the tertiary on the binary system. The inner mass-ratio of BD44 is 0.931 ( or 1/1.074) while for KIC65 it is 0.9383. The corresponding inner orbit periods of BD44 and KIC65 are 3.4726217d and 3.4205977d, respectively. The differences in the binaries of the two systems appear when we look at the LC solutions. The radius of the most massive star in BD44  is larger ($R_\mathrm{Ab}=2.4387_{-0.0048}^{+0.0035} R_\odot$) but it corresponds to the secondary (shallower) eclipse in the LC. This is also evident from the light-fractions obtained from the BF and also the fast rotation as reflected in the high $v\mathrm{sin}(i)$. Going by the convention, we will address this star as the secondary (BD44Ab; thus the secondary-to-primary mass ratio is $>$1). The primary of BD44 (BD44Aa) is inflated compared to the solar radius ($R_{Aa}=1.3118_{-0.0053}^{+0.0056} R_\odot$). Both the stars in KIC65 are very much solar-like ($R_{Aa}=1.0719_{-0.0020}^{+0.0019} R_\odot$ and $R_{Ab}=1.0101_{-0.0014}^{+0.0017} R_\odot$). A quick look at the MCMC corner plots (Fig. \ref{fig:ph_mcmc_BD}) tells us that the uncertainty on radii measurements is mostly affected by degeneracy in $L_\mathrm{Aa}$ which is itself degenerate with $l_3$. The $e_\mathrm{A}$ is adopted to be zero for both systems. But the MCMC maps show shifts of $2\sigma$ and $3\sigma$ from zero, for BD44 (Fig.\ref{fig:ph_mcmc_BD}) and KIC65 (Fig.\ref{fig:kicmcmc}) respectively. This small eccentricity can possibly be induced in the LC due to spots or be due to perturbations from the tertiary star. Unfortunately with current observations, it is difficult to decouple these effects.

 The temperatures of the BD44 binary stars are lower ($T^\mathrm{eff}_\mathrm{Aa}=5822\pm202 K$ and $T^\mathrm{eff}_\mathrm{Ab}=5449\pm100 K$ ) compared to those of KIC65 ($T^\mathrm{eff}_\mathrm{Aa}=6490\pm129 K$ and $T^\mathrm{eff}_\mathrm{Ab}=6397\pm123 K$). While the similar temperatures of KIC65Aa and KIC65Ab can explain the similar eclipse depths, we expect a temperature ratio of 0.772 (compared to 0.936 from the spectral analysis) from the LC fitting of BD44. This discrepancy can arise due to the cold spots in BD44, which affect the spectroscopic temperature measurement.
 
 The stars in BD44 have quite different $v\mathrm{sin}(i)$ with the BD44Aa having a value of 23.21$\, \mathrm{km\,s}^{-1}$ compared to 39.60$\, \mathrm{km\,s}^{-1}$ for BD44Ab. KIC65 has similar  $v\mathrm{sin}(i)$ for the stars in the inner binary as expected from stars with similar radii. But when comparing with calculated synchronised velocities, we find that the observed velocities are larger than expected. The  $v\mathrm{sin}(i)$ of the tertiary of KIC65 and BD44 are similar.  
 
\begin{figure}
    \centering
    \includegraphics[width=0.45\textwidth]{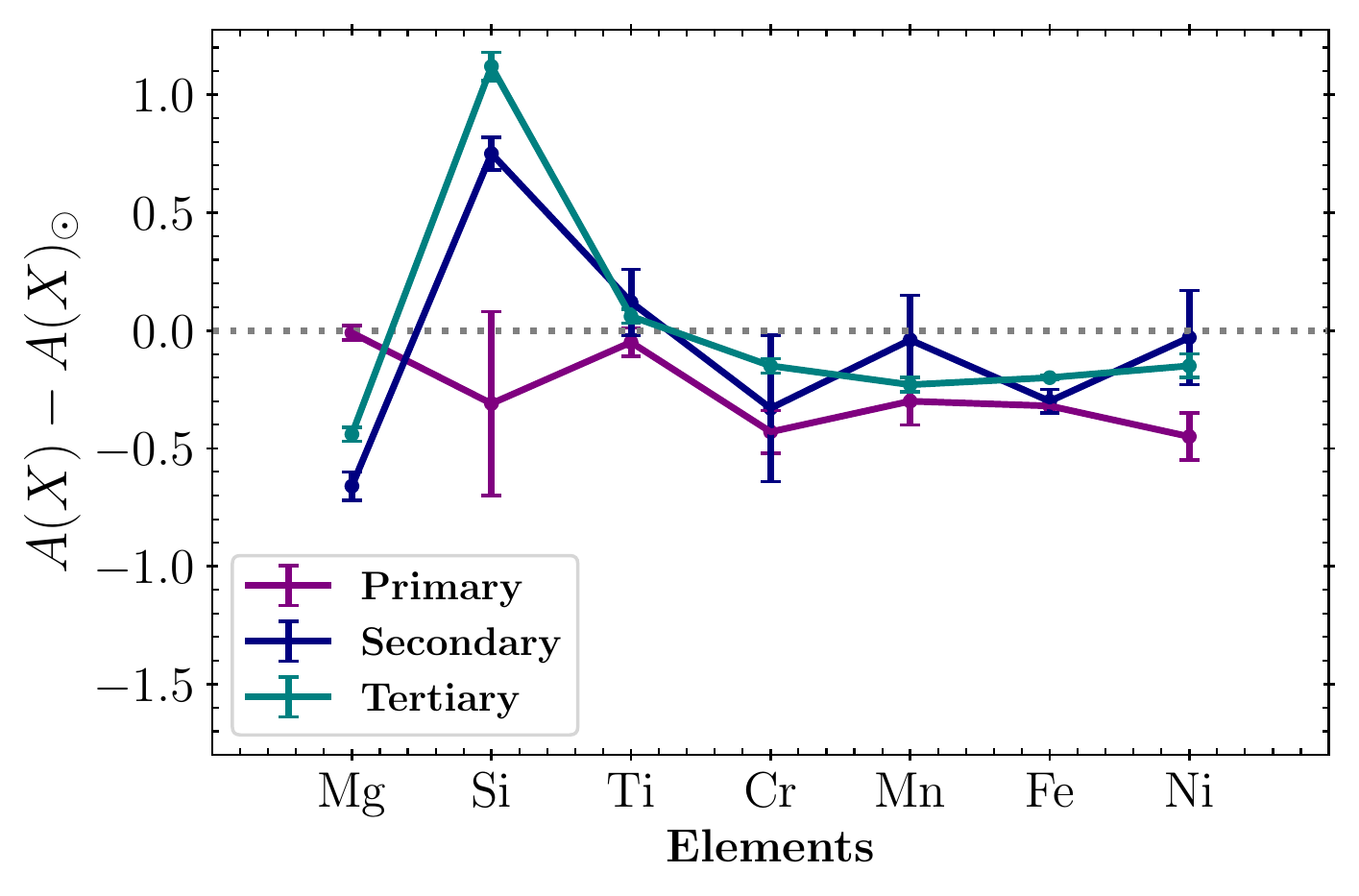}
        \includegraphics[width=0.45\textwidth]{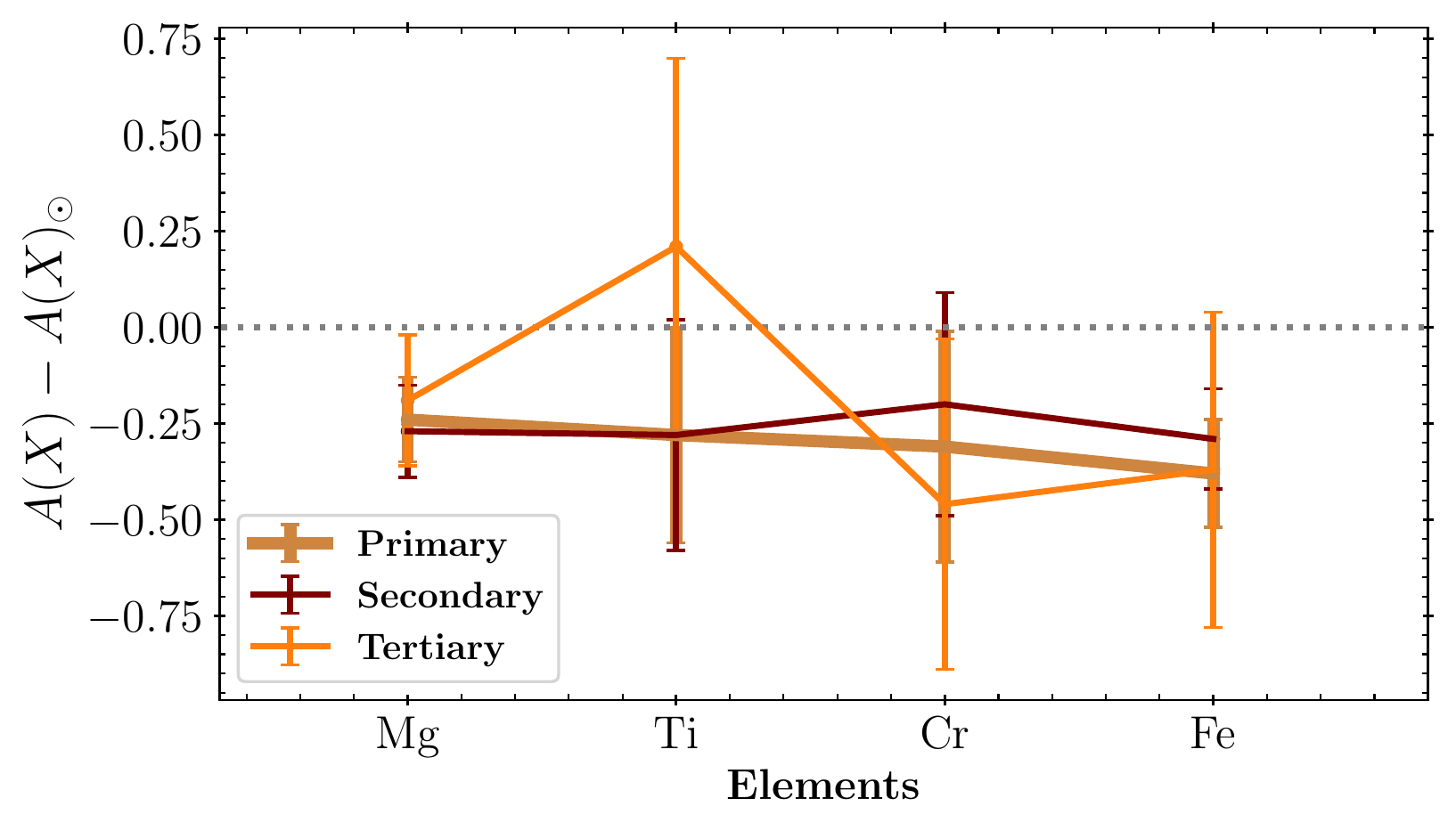}
    \caption{Abundance variation for all three stars in BD44 (top) and KIC65 (bottom). The plotted abundances are relative to solar abundances, which is denoted by the dotted line.}
    \label{fig:abunvar}
\end{figure}

  Both KIC65 and BD44 are metal-poor systems having [M/H] $-0.28\pm0.22$ and $-0.24\pm0.21$ respectively.  While most of the stars are found to be $\alpha$-enhanced systems, BD44Ab and BD44B were the only stars with negative $\alpha$. We calculated the abundances of some elements whose lines showed up on the spectra. We compare the abundances (see Tables \ref{tab:abunBD} and \ref{tab:abunKIC}) for all three stars, in both the systems, in Fig.\ref{fig:abunvar}. While all the stars in KIC65 have similar abundances (within error bars), BD44Ab and BD44B have a rise in $A(Si)$ and a dip in $A(Mg)$ compared to BD44Aa (Fig.\ref{fig:abunvar}). The total set of all parameters is given in Table. \ref{tab:paramtable} for comparison. 

\begin{table*}
 \centering
\caption{All adopted parameters for BD44 and KIC65 (except $i_m$). Unsymmetrical errors correspond to uncertainties estimated using MCMC sampling.  }
 \label{tab:paramtable}
 \small
\begin{tabular}{@{}lllllll}
\hline
 & \multicolumn{3}{c}{BD\,+44 2258} & \multicolumn{3}{c}{KIC\,06525196} \\
\hline
\multicolumn{7}{c}{Orbital Parameters} \\
\hline
   & \multicolumn{2}{c}{Aa--Ab} & A--B  & \multicolumn{2}{c}{Aa--Ab} & A--B  \\
  \hline
  $t_0$ [BJD - 2450000]& \multicolumn{3}{c}{$8740.8268426$} & \multicolumn{3}{c}{$2421.6563579$} \\
  $P$ [days] & \multicolumn{2}{c}{$3.4726217\pm0.0000015$} & $254.84\pm0.05$  & \multicolumn{2}{c}{$3.42059774\pm0.00000014$} & $418.0\pm0.4$ \\
  \vspace{0.1cm}
  $a$ [R$_\odot$] & \multicolumn{2}{c}{$12.06\pm0.12$} & $240.2\pm3.3$ & \multicolumn{2}{c}{$12.053\pm0.018$} & $330.8\pm2.1$ \\
  \vspace{0.1cm}
  $e$ & \multicolumn{2}{c}{$0.000^a$} & $0.598\pm0.002$ & \multicolumn{2}{c}{$0.000^a$} & $0.301\pm0.003$ \\
  $i$ [deg] & \multicolumn{2}{c}{$80.1271_{-0.0057}^{+0.0084}$} & $76.15_{-1.61}^{+1.91}$ & \multicolumn{2}{c}{$84.7083_{-0.0063}^{+0.0051}$} & $84.7_{-2.6}^{+5.3}$ \\
  $\omega$ [deg]& \multicolumn{2}{c}{$-$} & $203.0\pm0.2$ & \multicolumn{2}{c}{$-$} & $276\pm1$ \\
  $q$ & \multicolumn{2}{c}{$1.074\pm0.021$} & $0.465\pm0.019$ & \multicolumn{2}{c}{$0.9383\pm0.0026$} & $0.3875\pm0.0045$ \\
  $K_1$ [km\,s$^{-1}$] & \multicolumn{2}{c}{$89.66\pm0.98$} & $18.30\pm0.75$ & \multicolumn{2}{c}{$85.96\pm0.12$} & $11.67\pm0.12$ \\ 
  $K_2$ [km\,s$^{-1}$] & \multicolumn{2}{c}{$83.47\pm1.37$} & $39.39\pm0.22$ & \multicolumn{2}{c}{$91.62\pm0.12$} & $30.15\pm0.17$ \\ 
 \hline
\multicolumn{7}{c}{Stellar and atmospheric parameters} \\
\hline
   & Aa & Ab &  B & Aa & Ab &  B \\
 Flux fraction (from Spectroscopy) & $0.3758\pm0.0268$ & $0.5139\pm0.0285$ & $0.1104\pm0.0088$ & $0.4851\pm0.0225$ & $0.4010\pm0.0154$ & $0.1139\pm0.0181$ \\
 \vspace{0.1cm}
 Flux fraction (from Photometry) & $0.4021_{-0.0030}^{+0.0039}$ & $0.5101_{-0.0031}^{+0.0021}$ & $0.0878_{-0.0015}^{+0.0016}$  & $0.4990_{-0.0084}^{+0.0101}$ & $0.4060_{-0.0068}^{+0.0051}$ & $0.0950_{-0.0034}^{+0.0033}$ \\
 \vspace{0.1cm}
 $M$ [M$_\odot$] & $0.941\pm0.033$ & $1.011\pm0.029$ & $0.907\pm0.065$ & $1.0351\pm0.0055$ & $0.9712\pm0.0039$ & $0.777\pm0.012$\\
 \vspace{0.1cm}
 $R$ [R$_\odot$] & $1.3118_{-0.0053}^{+0.0056}$ & $2.4387_{-0.0048}^{+0.0035}$ & $1.67_{-0.59}^{+0.70}$ & $1.0719_{-0.0019}^{+0.0020}$ & $1.0101_{-0.0017}^{+0.0014}$ & $0.74_{-0.38}^{+0.80}$ \\
 
 $T_\mathrm{eff}$ [K]& $5822\pm202$ & $5449\pm100$ & $5261\pm181$ & $6490\pm129$ & $6397\pm123$ & $5393\pm319$ \\
 $\log(g)$ [dex] & $4.19^b$ & $3.69^b$ & $3.95\pm0.39$ & $4.39^b$ & $4.41^b$ & $4.58\pm0.63$ \\
 $v_\mathrm{mic}$ [km~s$^{-1}$] & $2.36\pm0.75$ & $1.78^c$ & $1.19\pm0.5$ & $1.39\pm0.54$ & $1.18\pm0.53$ & $1.5^c$ \\
 $v_\mathrm{mac}$ [km~s$^{-1}$]$^{c}$ & $4.69$ & $3.76$ & $3.43$ & $9.28$ & $8.43$ & $5.94$ \\
 $v\mathrm{sin}(i)$ [km~s$^{-1}$] & $23.21\pm 2.19$ & $39.60\pm3.26$ & $5.34\pm1.41$ & $20.69\pm1.45$ & $18.39\pm1.15$ & $5.23\pm1.34$ \\
 $\alpha$  [dex]    & $0.21 \pm 0.12$ & $-0.22 \pm 0.08$ & $-0.21 \pm 0.11$ & $0.07 \pm 0.11 $ &$0.16 \pm 0.11$ & $0.29 \pm 0.24$ \\
 \hline
\multicolumn{7}{c}{System parameters} \\
  \hline
  \vspace{0.1cm}
$\log$(age) [dex] &\multicolumn{3}{c}{$9.89_{-0.05}^{+0.03}$} & \multicolumn{3}{c}{$9.49\pm0.06$} \\
  \vspace{0.1cm}
$\mathrm{[M/H]}_\mathrm{iSpec}$  [dex]   &\multicolumn{3}{c}{$-0.24\pm0.21$} & \multicolumn{3}{c}{$-0.28\pm0.22$} \\
  \vspace{0.1cm}
$\mathrm{[Fe/H]}_\mathrm{isoc}$  [dex]   &\multicolumn{3}{c}{$-0.40^{+0.15}_{-0.10}$} & \multicolumn{3}{c}{$-0.45^{+0.15}_{-0.35}$} \\
\vspace{0.1cm}
$E(B-V)^{d}$ [mag]    &\multicolumn{3}{c}{$0.176^{+0.019}_{-0.049}$} & \multicolumn{3}{c}{$0.053^{+0.040}_{-0.030}$} \\
Distance$^{d}$ [pc]   &\multicolumn{3}{c}{$194.5_{-10.3}^{+3.3}$} & \multicolumn{3}{c}{$220.01^{+2.26}_{-1.68}$} \\  
\hline
\end{tabular}

$^a$ Fixed while optimisation. $^b$ Fixed from LC fitting solutions. $^c$ Obtained from empirical tables. $^d$ Based on isochrone fitting. 
\end{table*}

\begin{table}
\begin{center}

\caption[]{Abundances of individual elements of the components of BD44. Solar abundances are presented for comparison (\citealt{2009ARA&A..47..481A}).}\label{BDabunresult}
 \label{tab:abunBD}
 \begin{tabular}{lccccc}
  \hline\noalign{\smallskip}
  Elements & Primary  & Secondary   & Tertiary  & Solar             \\
  \hline\noalign{\smallskip}
  
$_{12}$Mg &7.59\,$\pm$\,0.03  & 6.94\,$\pm$\,0.06  & 7.16\,$\pm$\,0.03& 7.60\,$\pm$\,0.04\\
$_{14}$Si &7.20\,$\pm$\,0.39  & 8.26\,$\pm$\,0.07  & 8.63\,$\pm$\,0.06& 7.51\,$\pm$\,0.03\\
$_{22}$Ti &4.90\,$\pm$\,0.06 & 5.25\,$\pm$\,0.14 &   5.01\,$\pm$\,0.03  & 4.95\,$\pm$\,0.05\\
$_{24}$Cr &5.21\,$\pm$\,0.09 & 5.31\,$\pm$\,0.31  &  5.49\,$\pm$\,0.03   & 5.64\,$\pm$\,0.04\\
$_{25}$Mn &5.13\,$\pm$\,0.10  & 5.39\,$\pm$\,0.19  &   5.20\,$\pm$\,0.03  & 5.43\,$\pm$\,0.05			   \\
$_{26}$Fe &7.18\,$\pm$\,0.03 & 7.20\,$\pm$\,0.05 &   7.30\,$\pm$\,0.01  & 7.50\,$\pm$\,0.04\\
$_{28}$Ni &5.77\,$\pm$\,0.10 & 6.19\,$\pm$\,0.20 &   6.07\,$\pm$\,0.05  & 6.22\,$\pm$\,0.04\\
  \noalign{\smallskip}\hline
\end{tabular}
\end{center}  
\end{table}

\begin{table}

\begin{center}
\caption[]{Abundances of individual elements of the components of KIC65.}\label{KICabunresult}
 \label{tab:abunKIC}
 \begin{tabular}{lccccc}
  \hline\noalign{\smallskip}
  Elements & Primary  & Secondary   & Tertiary  & Solar             \\
  \hline\noalign{\smallskip}
$_{12}$Mg & 7.36 \,$\pm$\, 0.11 & 7.33 \,$\pm$\, 0.12 & 7.41 \,$\pm$\, 0.17 & 7.60\,$\pm$\,0.04\\
$_{22}$Ti & 4.67 \,$\pm$\, 0.28 & 4.67 \,$\pm$\, 0.30 & 5.16 \,$\pm$\, 0.49 & 4.95\,$\pm$\,0.05\\
$_{24}$Cr & 5.33 \,$\pm$\, 0.30 & 5.44 \,$\pm$\, 0.29 & 5.17 \,$\pm$\, 0.43 & 5.64\,$\pm$\,0.04\\
$_{26}$Fe & 7.12 \,$\pm$\, 0.14 & 7.21 \,$\pm$\, 0.13 & 7.13 \,$\pm$\, 0.41 & 7.50\,$\pm$\,0.04\\
 \noalign{\smallskip}\hline
\end{tabular}
\end{center}  
\end{table}

 \subsection{Physical parameters of the Tertiary}
\begin{figure}
    \centering
    \includegraphics[width=0.45\textwidth]{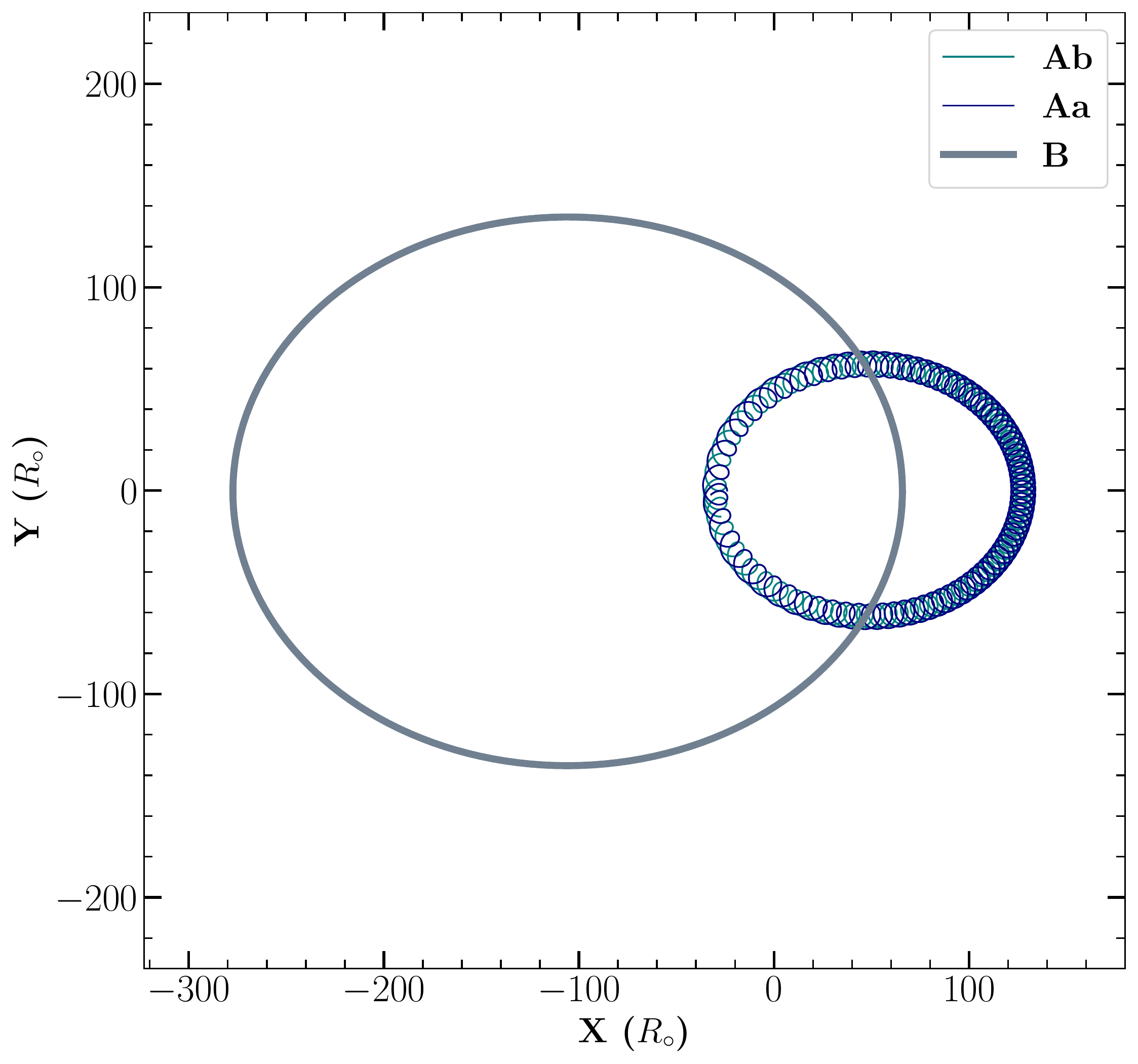}
    \includegraphics[width=0.45\textwidth]{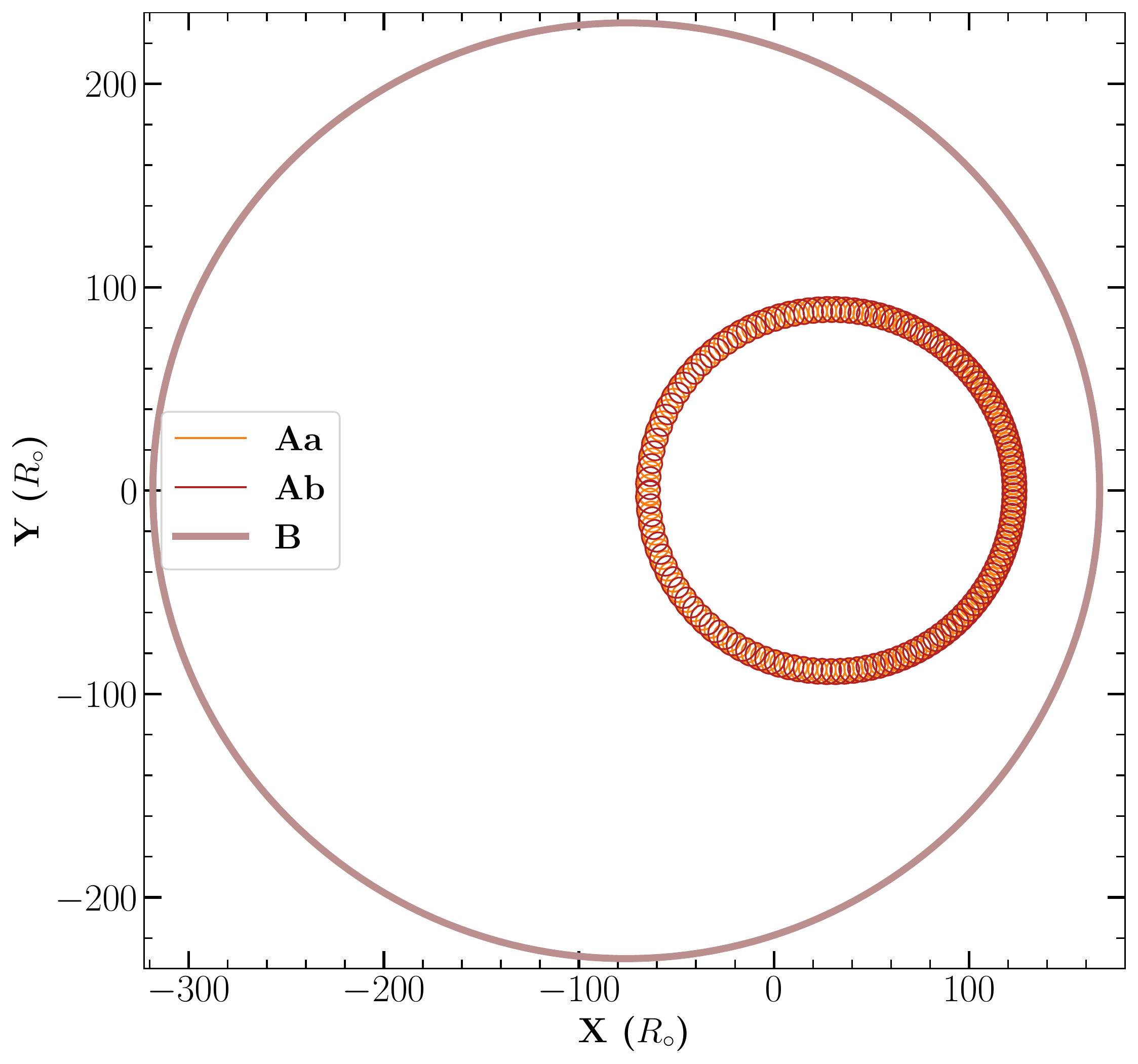}
    \caption{Orbital path of all the stars of BD44 (top) and KIC65 (bottom). The orbits are viewed perpendicular to the orbital plane of the inner binary (for $i_{m}=4^{\circ}$) and are integrated over one outer-orbit period.}
    \label{fig:orbitsectionBD}
\end{figure}

The tertiary stars orbit the inner binary with the periods of $254.84\pm0.05$  and  $418.0\pm0.4$ d for BD44 and KIC65, respectively.
They are different in both mass and radius. The mass of the tertiary in BD44 is close to solar ($0.907\pm0.065$\,M$_\odot$) while tertiary of KIC65 is less massive ($0.777\pm0.012$\,M$_\odot$). The estimated radii have large errors, but a comprehensive look at all the signatures suggests that the tertiary in BD44 has an inflated atmosphere and therefore a radius larger than one expected for a main-sequence star of this mass. The tertiary of KIC65 is most likely to have a radius of ~0.74 $R_\odot$. The stars themselves are orbiting with different periods around the inner binary system. The tertiary of BD44 is in an orbit with higher eccentricity ($e_\mathrm{AB} = 0.598$) than KIC65 ($e_\mathrm{AB}\sim 0.3$;  Fig.\ref{fig:orbitsectionBD}). These different configurations will mostly affect the timescale of secular perturbations which depend on $e_\mathrm{AB}$ \citep{fordseculartrip}. 

\subsection{Possible Mutual Inclinations}
\begin{figure}
    \centering
    \includegraphics[width=0.45\textwidth]{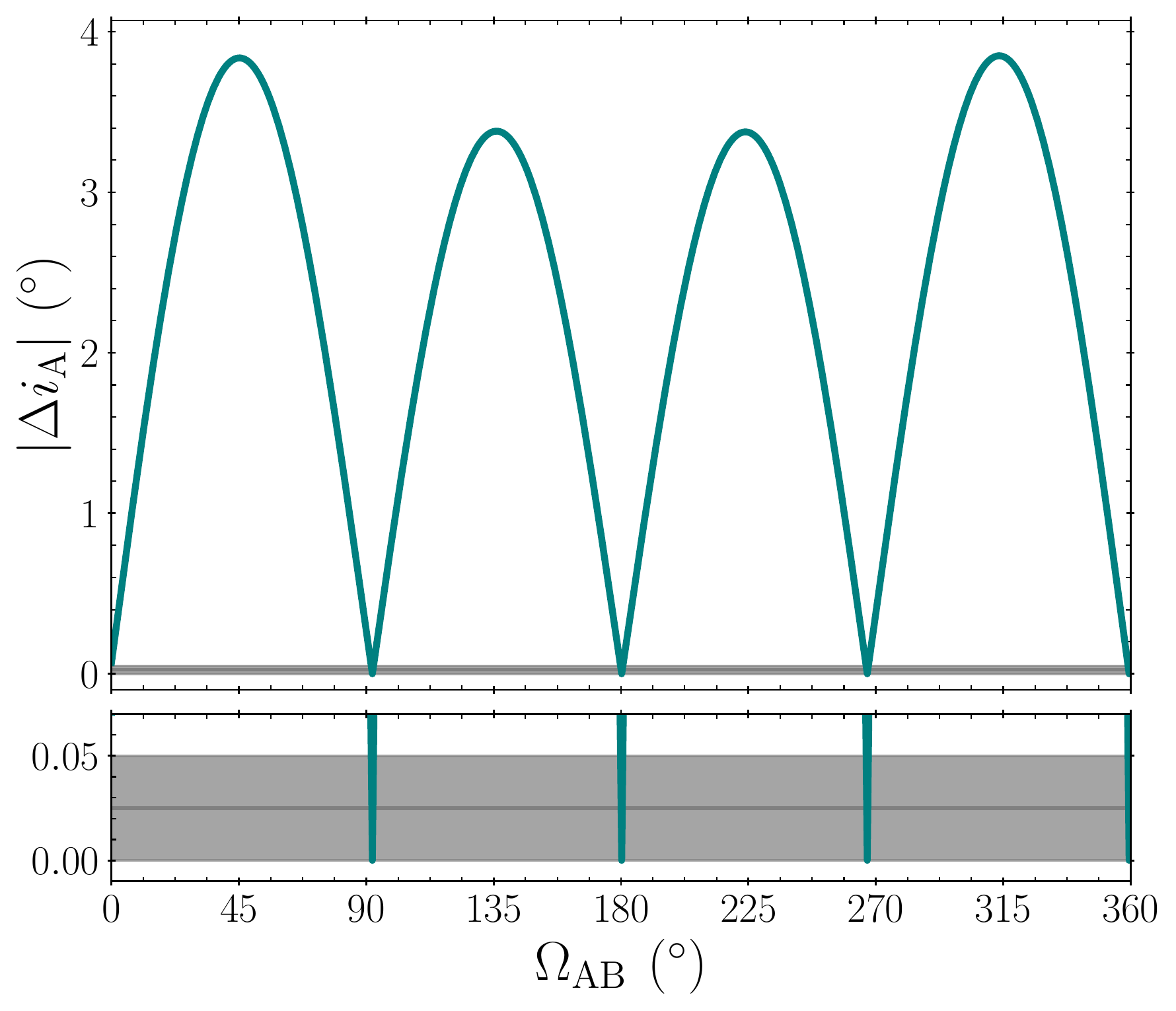}
    \includegraphics[width=0.45\textwidth]{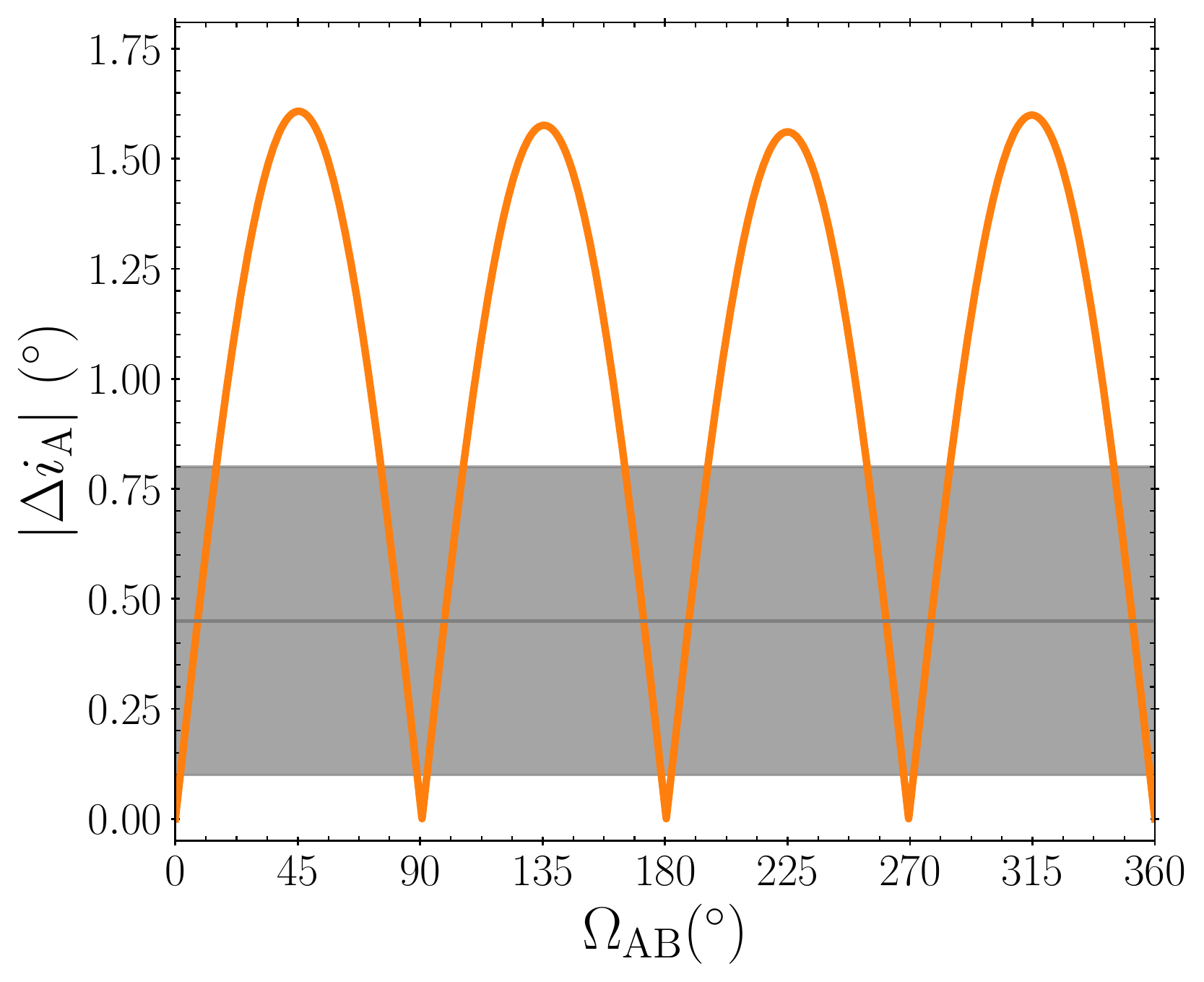}
    \caption{Variation of $|\Delta i_\mathrm{A}|$ from numerical simulations for different values of $\Omega_\mathrm{AB}$, for BD44 (top) and KIC65 (bottom). The timescale of the simulation corresponds to the length of the available LCs. The grey lines show the observed $|\Delta i_\mathrm{A}|$ with the shaded regions representing the errors in measurements. The overlapping regions represent the possible values of $\Omega_\mathrm{AB}$. The lower panel of the top figure represents a zoomed-in view of the observed variations.}
    \label{fig:omegavsi}
\end{figure}

Short-term numerical integration gave us an estimate of inclination changes of the inner binary ($\Delta i_\mathrm{A}$) for different values of $\Omega_\mathrm{AB}$. We then obtained observed $\Delta i_\mathrm{A}$ by using inclination values obtained by \cite{krishides2017} for KIC65, subtracted from the values obtained in this work. While for BD44 we used the inclinations observed in Sector-16 and Sector-49 of \TESS{} LCs (Table.\ref{tab:incvar}). This gave us possible values of $\Omega_\mathrm{AB}$  for the observed $\Delta i_\mathrm{A}$ (Fig.\ref{fig:omegavsi}). We then used Eq.(4) to translate the possible $\Omega_\mathrm{AB}$ values to possible $i_m$ values for all possible configurations (Table.\ref{tab:mutinc}).

\begin{table}

\begin{center}
\caption[]{Variation of $i_\mathrm{A}$ over time.}
 \label{tab:incvar}
 \begin{tabular}{cccc}
\hline
\hline
  System & Initial $i_\mathrm{A}$ [deg]& Final $i_\mathrm{A}$ [deg]&  Time (yrs)  \\ 
  \hline
    BD44 & $80.1271 \pm 0.008$ & $80.1521 \pm 0.017$ & 2.7 \\ 
    KIC65 & $85.15 \pm 0.34$ & $84.7083 \pm 0.005$ & 6.9  \\
\hline
\hline
\end{tabular}
\end{center}  
\end{table}

\begin{table}

\begin{center}
\caption[]{Estimates of mutual inclination for possible values of $\Omega_\mathrm{AB}$.}
 \label{tab:mutinc}
 \begin{tabular}{cccc}
  \hline\noalign{\smallskip}
  System & $\Omega_\mathrm{AB}$&\multicolumn{2}{c}{$i_m$ }         \\  
  \hline\noalign{\smallskip}
  & & Config.A & Config.B \\
  \hline
  \hline\noalign{\smallskip}
  \multirow{4}{*}{BD44}& $359.46 \pm 0.76$ & $4.03 \pm 0.10$ & $175.97 \pm 0.01$   \\
  & $266.97 \pm 0.79$ & $90.56 \pm 0.76$ &  $89.45 \pm 0.76$ \\
  & $180.23 \pm 0.83$ &  $156.27 \pm 0.01$&  $23.73 \pm 0.01$\\
  & $ 92.16 \pm 0.80$& $89.71 \pm 0.77$& $90.29 \pm 0.77$\\
  \hline
  \multirow{4}{*}{KIC65}& $353.80 \pm 4.85$ & $6.18 \pm 4.83$ & $173.82 \pm 4.83$   \\
  & $269.41 \pm 11.12$ & $90.09 \pm 11.02$ &  $89.91 \pm 11.02$ \\
  & $180.42 \pm 11.18$ &  $164.60 \pm 0.30$&  $15.40 \pm 0.30$\\
  & $ 90.67 \pm 11.10$& $90.16 \pm 11.00$& $89.83 \pm 11.00$\\
  & $6.19 \pm  4.85$  & $6.17 \pm 4.83$ & $173.83 \pm 4.83$ \\
\hline
\hline
\end{tabular}
\end{center}  
\end{table}

\subsection{Age and Evolution}
\begin{figure}
    \centering
        \includegraphics[width=0.45\columnwidth]{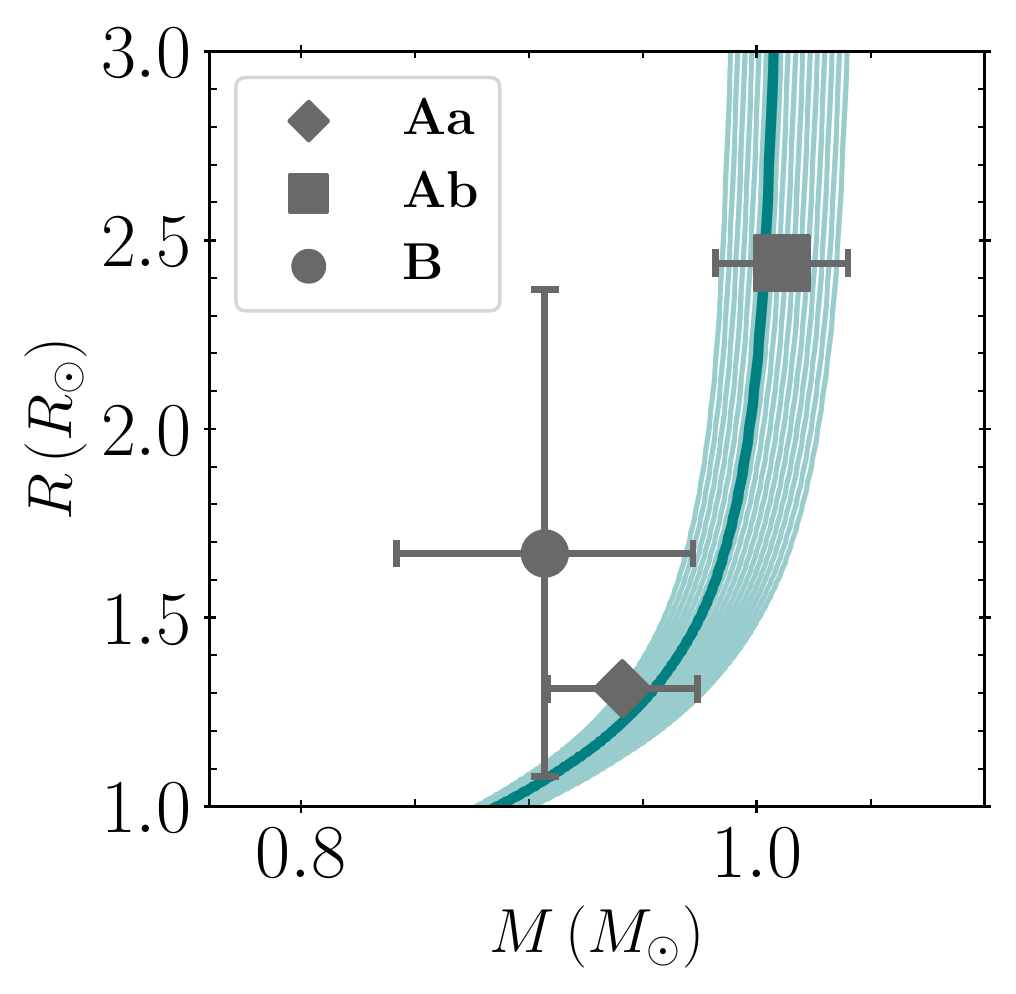}
    \includegraphics[width=0.465\columnwidth]{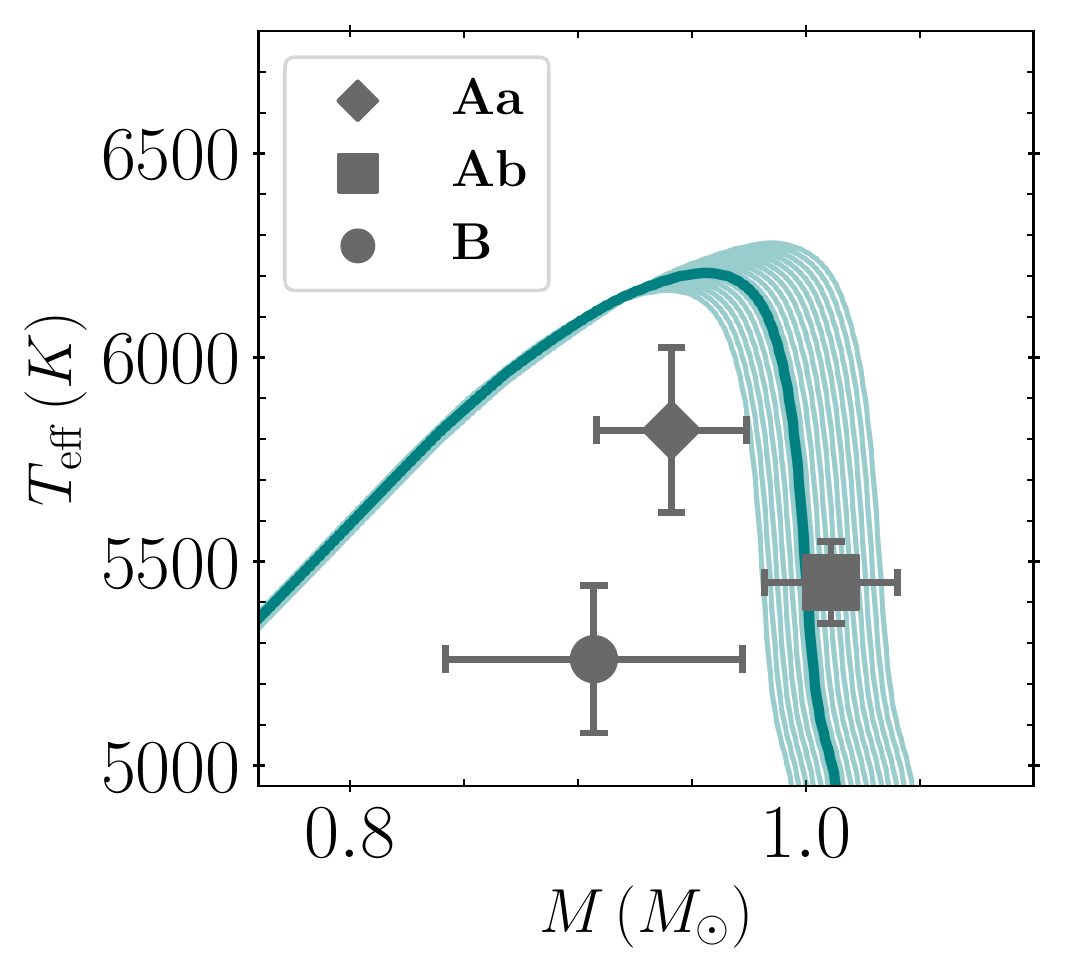} \\
    \includegraphics[width=0.45\columnwidth]{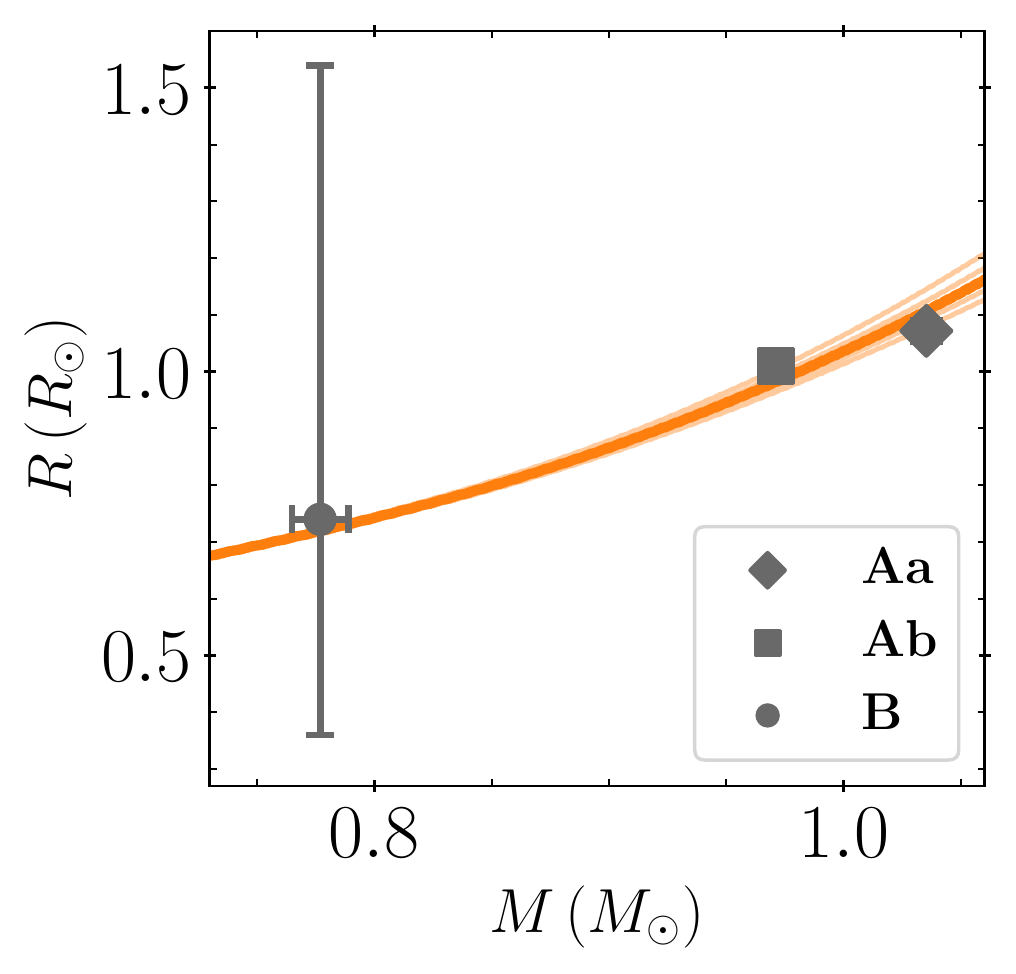}
    \includegraphics[width=0.465\columnwidth]{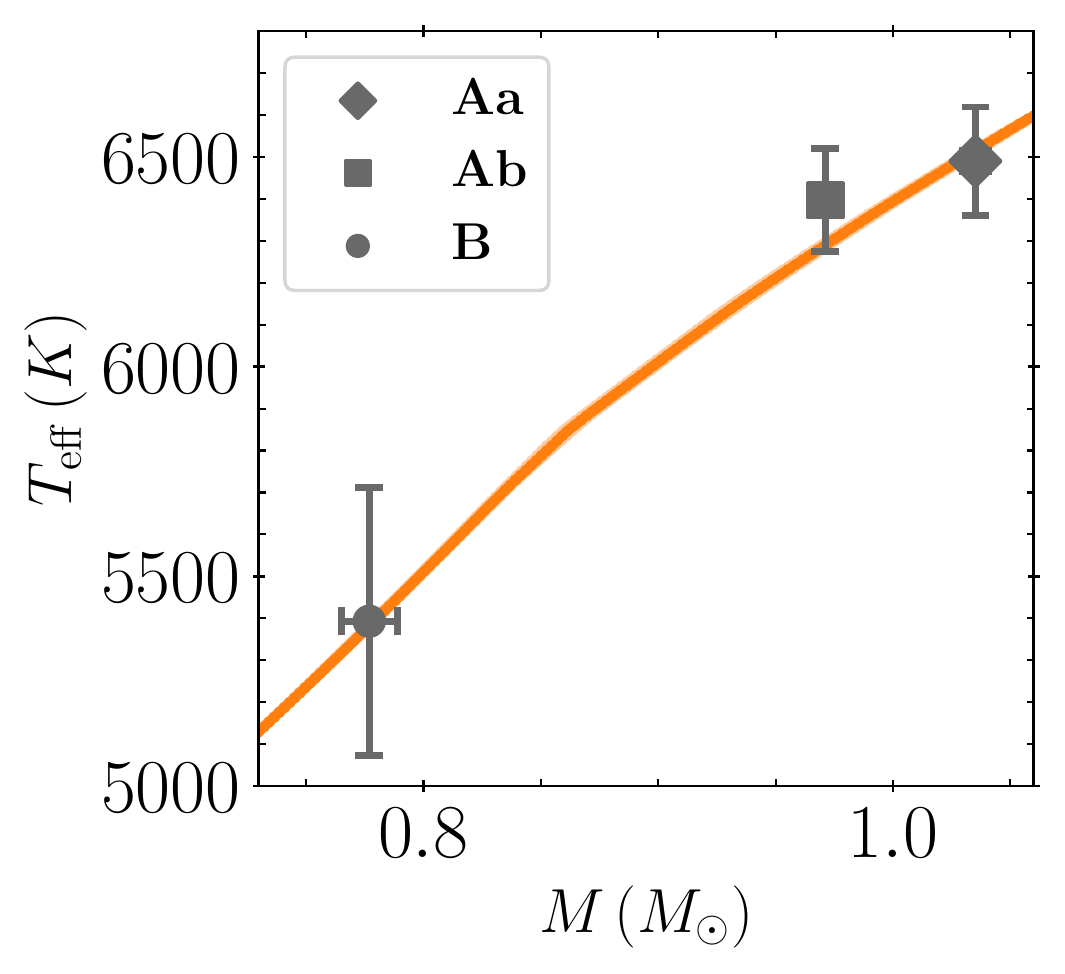}
    \caption{Mass-Radius (left column) and Mass-Temperature (right column) isochrone fits for BD44 (top; in green) and KIC65 (bottom; in orange). The shaded area represents the parameter-space corresponding to the errors of the fitting. The respective measurements from the stars are plotted in grey. The size of the points are relative to their radii.}
    \label{fig:isofit}
\end{figure}

Isochrone fitting puts the log(age) for BD44 between 9.84 and 9.92 (95\% confidence level), for metallicity range of -0.25 to -0.50 dex. The formally best fit was found for log(age) = 9.89 (7.8~Gyr) and [Fe/H] = -0.40~dex. While the primary of BD44, BD44Aa, is a main-sequence star, BD44Ab is a sub-giant. With the large uncertainties in the parameters of BD44B, it is hard to determine its evolutionary state. But the simultaneous mass-radius and mass-temperature isochrone fit depicts it as a sub-giant star (Fig.\ref{fig:isofit}). The other signatures of BD44B being a sub-giant are found in (i) large amplitude of BF (Fig.\ref{fig:BFunction}), and (ii) similar abundances (Fig.\ref{fig:abunvar}) as that of the BD44Ab (which is a sub-giant itself). However, with the available data, we can not completely rule out the possibility that it is a main-sequence star, less massive and smaller than the primary.

The isochrone-based, reddening-free distance was found to be 194.5~pc, which is significantly larger than the GDR3 Part~3 value of 165.8$\pm$0.6~pc. The distribution of acceptable models is very skewed, and none of the acceptable models reached a distance lower than 184~pc. The tension is probably caused by the tertiary, which was formally found to be less massive but seemingly more evolved than the primary. Its parameters could probably be better determined with additional observations around the outer orbit's pericenter, where the tertiary's RVs reach their minimum. It should also be noted that the GDR3 solution is of worse quality than for KIC65.

Even though having the binary stars with masses similar to BD44, both the primary and secondary of KIC65 are main-sequence stars along with its tertiary. This immediately suggests that KIC65 is significantly younger than BD44. The fitting procedure for KIC65 resulted in log(age) = 9.49 (3.1~Gyr) and [Fe/H] = $-0.45$~dex, with the 95\% confidence level ranges of 9.43 to 9.55, and $-0.80$ to $-0.30$~dex, respectively. The isochrone-based reddening-free distance ($\sim$220$\pm$2~pc) is in excellent agreement with, and of comparable precision to, the GDR3 solution for an astrometric binary model (222.3$\pm$1.7~pc), even when it was not used as a constraint.

\subsection{Dynamical Evolution}
\begin{figure}
    \centering
    \includegraphics[width=0.45\textwidth]{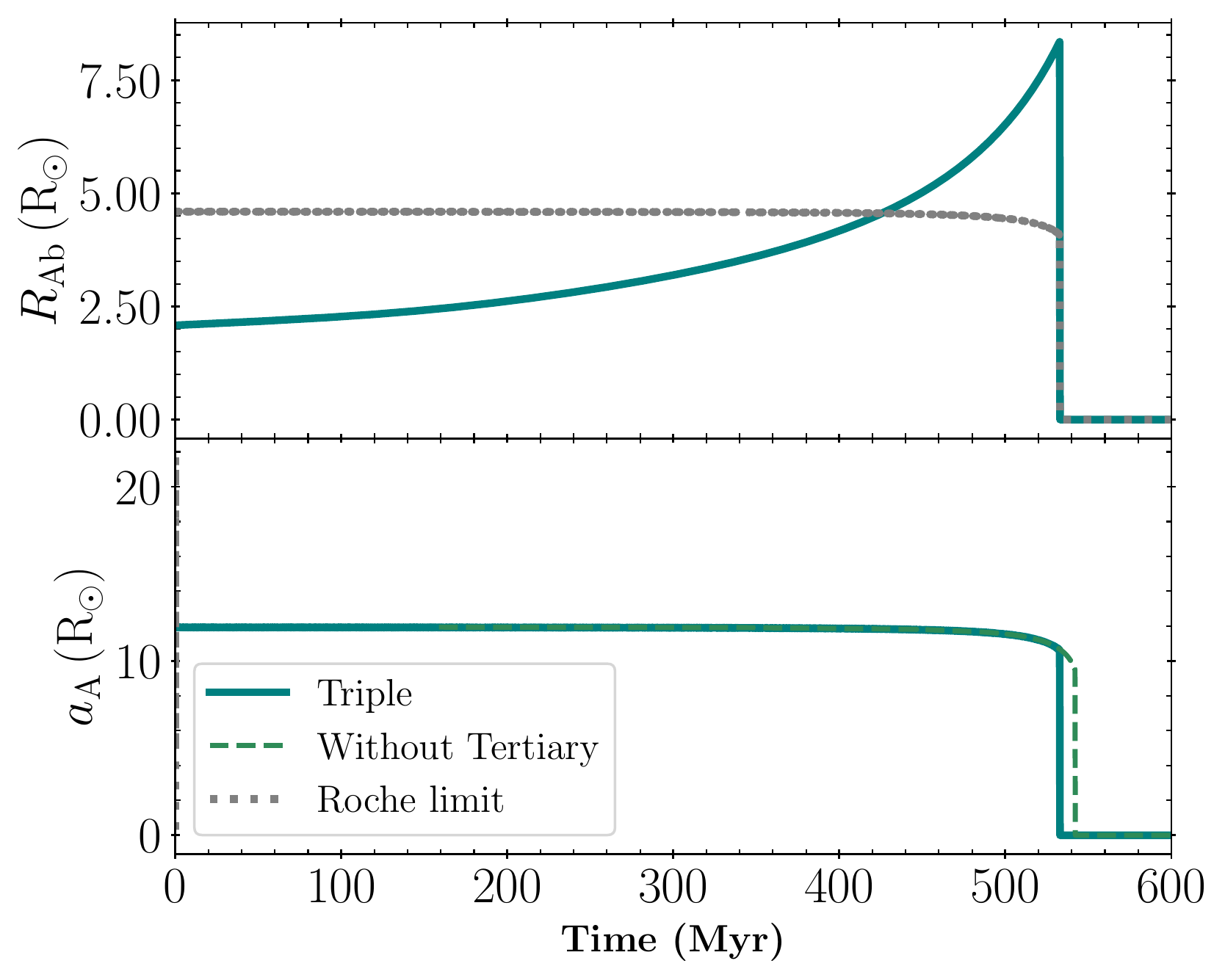}
    \caption{Variation of the radius of BD44Ab and the semi-major axis of the inner binary of BD44 as simulated in \textsc{reboundx} for $i_{m}=4^{\circ}$. The radius variation is interpolated from MIST grids. The grey dotted line in the top panel shows Roche-limit for the system. The dashed line in the lower panel shows the change in the collision time if no tertiary star was present.}
    \label{fig:BD44dynevo}
\end{figure}

\begin{figure}
    \centering
    \includegraphics[width=0.45\textwidth]{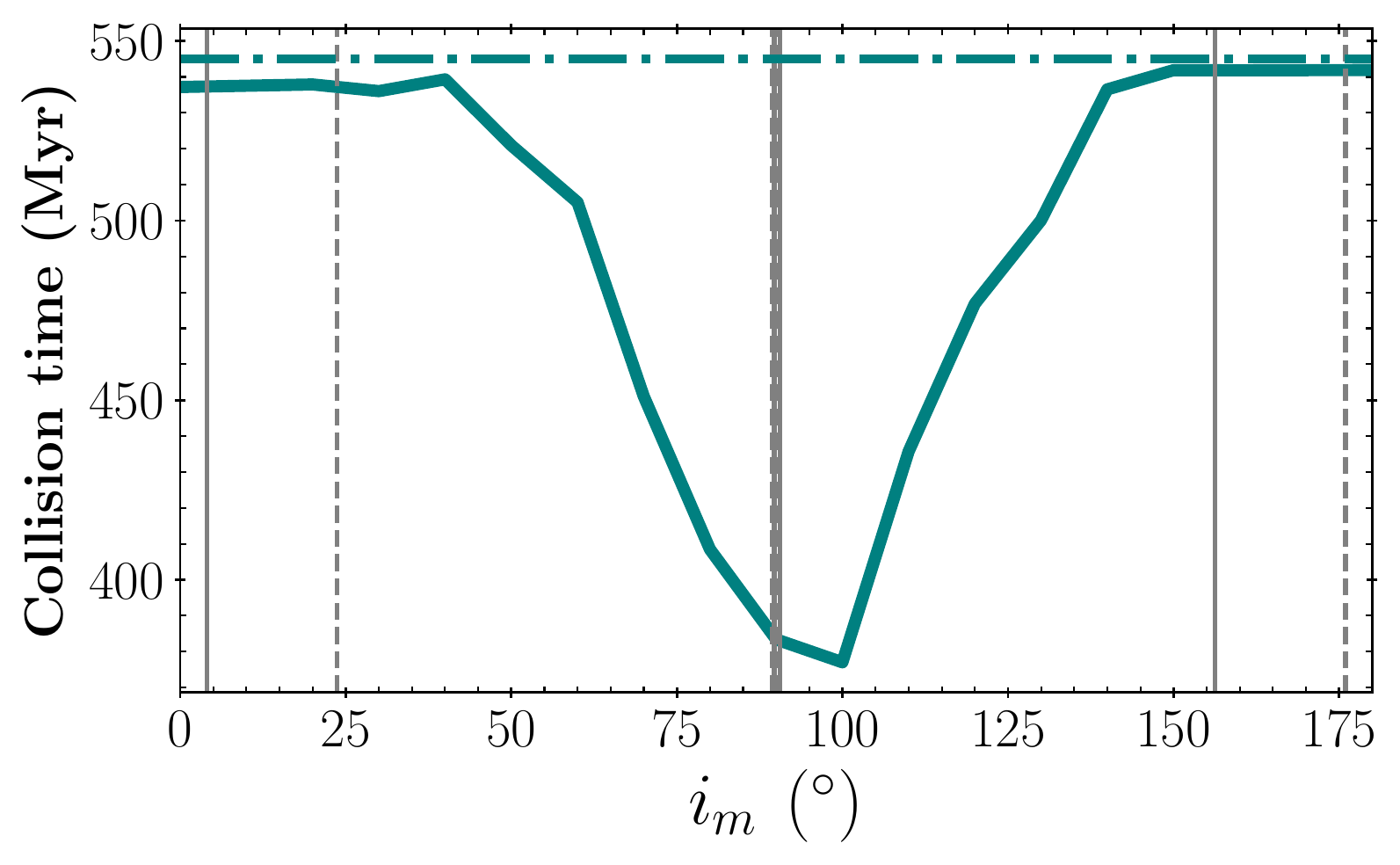}
    \caption{The simulated collision time of the inner binary of BD44 for different mutual inclinations. The green dot-dashed line shows the time for collision without a tertiary companion. The vertical grey lines represent the possible values of $i_m$. Solid lines represent possible $i_m$ for Config.A while dashed lines are for Config.B.}
    \label{fig:collisiondist}
\end{figure}

Long-term evolution of KIC65 shows that the system is stable for 600 Myrs more. But BD44 becomes a binary system within 550 Myrs due to the collision/merger of the inner binary. This collision will be driven by increasing tidal forces due to the increasing radius of the sub-giant BD44Ab (Fig.\ref{fig:BD44dynevo}: upper-panel). The radius of the star will exceed the Roche-limit \citep{eggletonrocheradii} at around 450 Myr and will drive the merger process unless the formation of a contact binary stabilises this system. This merger is mostly due to the tides in the inner binary as a lack of tertiary companion would have only delayed the merger by a few Myrs (Fig.\ref{fig:BD44dynevo}: lower-panel) for most of the estimated $i_m$. But a $i_m$  near $90^{\circ}$ will make the system merge faster ($<$400~Myr) than the lower values of $i_m$ (Fig.\ref{fig:collisiondist}).

\subsection{Spot Evolution}

\begin{figure}
    \centering
    \includegraphics[width=0.8\columnwidth]{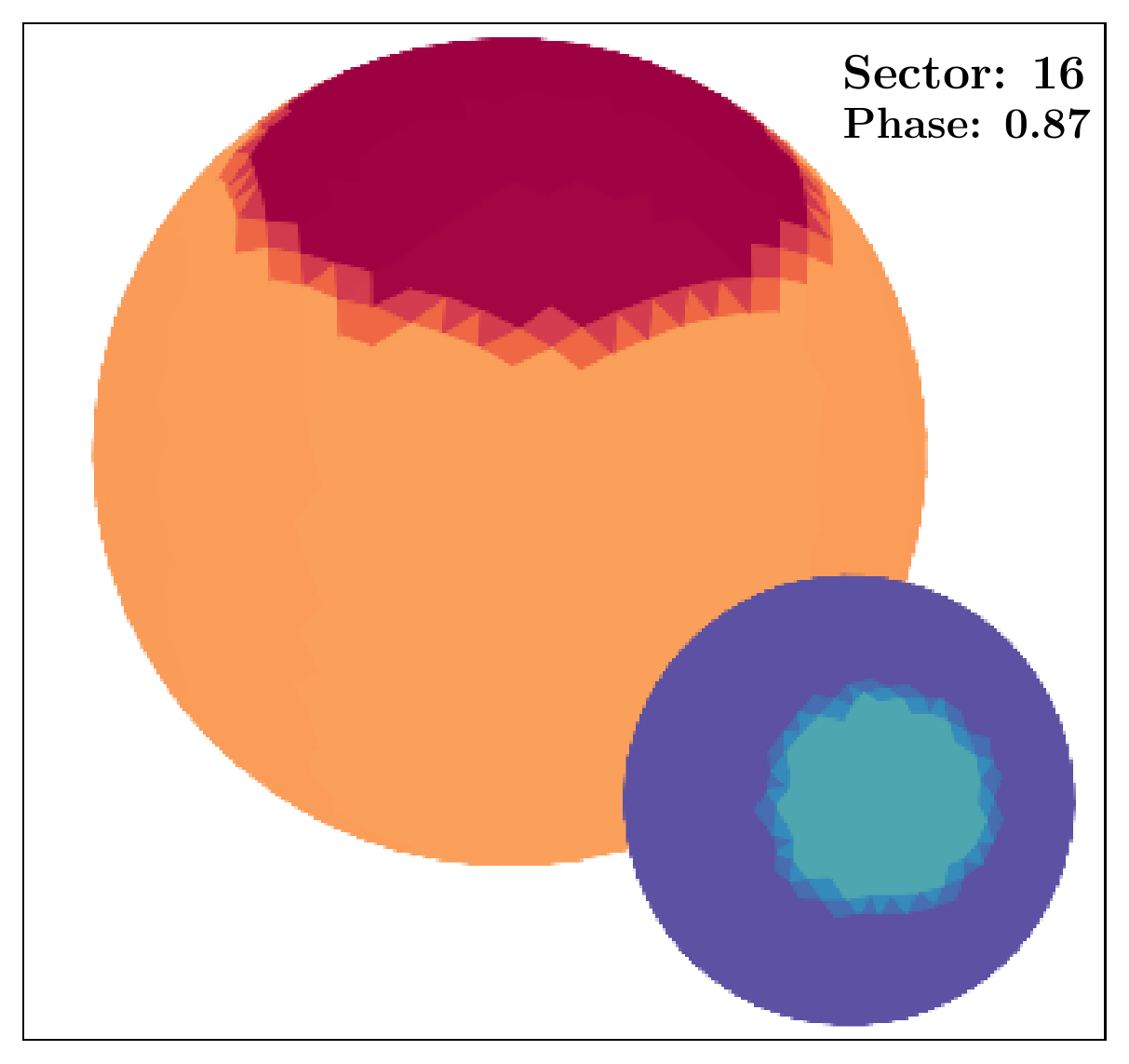}
        \includegraphics[width=0.8\columnwidth]{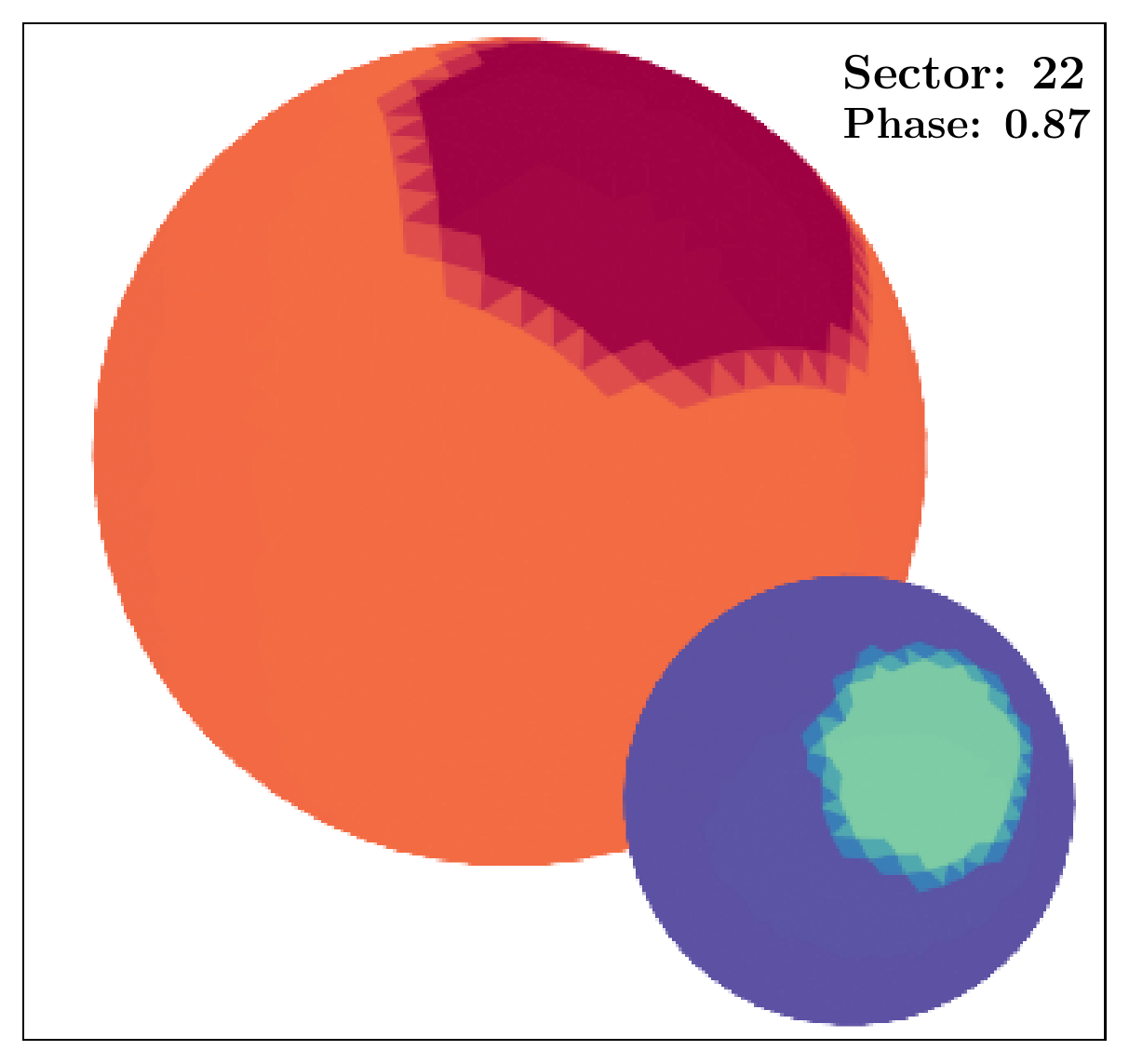}
         \includegraphics[width=0.8\columnwidth]{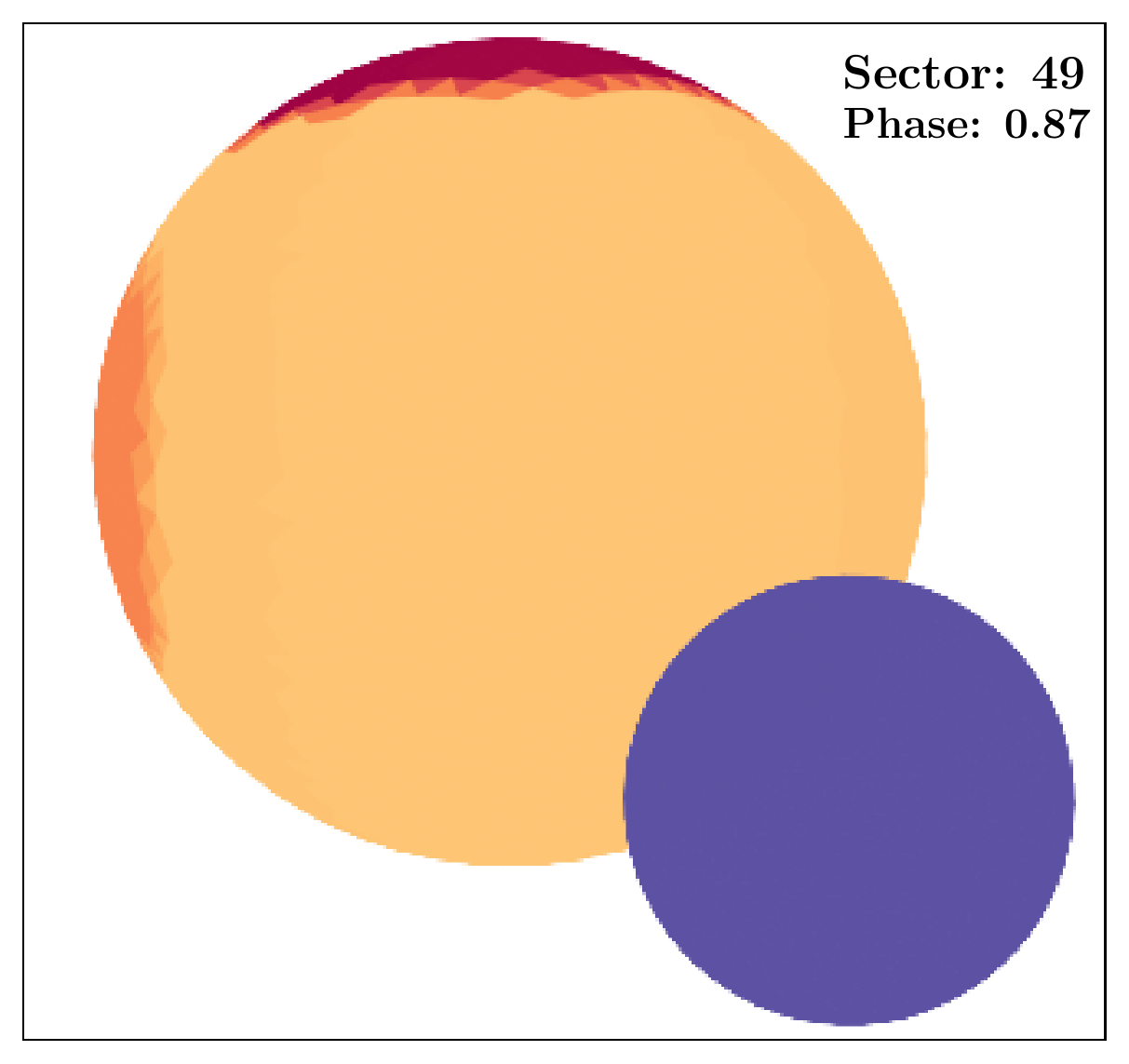}
         \includegraphics[width=0.8\columnwidth]{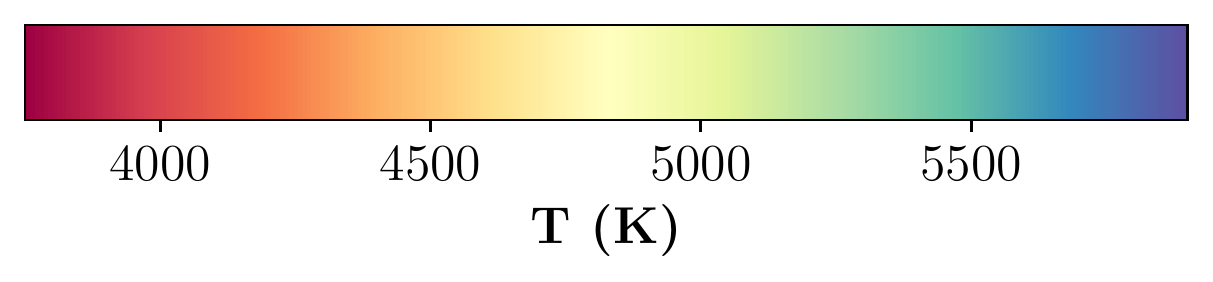}
    \caption{Spot positions on the inner binary stars of BD44 (at the same phase) for Sector-16 (top), Sector-22 (middle), and Sector-40 (bottom) as obtained with \PHOEBE{} modelling. The stable-spot near the pole of the secondary is the spot that affects the eclipse depths the most and is the one that migrates with a timescale of $\sim$ 5 yr.}
    \label{fig:spots}
\end{figure}

The light curve of BD44 is highly varying over different sectors owing to the migrating and evolving cold spots. The activity of BD44 is corroborated by its ultraviolet and X-ray emissions.  The distortions on the BFs of the stars in BD44 indicate the secondary (BD44Ab) to have more spots.  The occurrence of the fast-evolving and migrating spots is common on sub-giant stars. The biggest spot on the secondary is still visible in the newest Sector-49 of \TESS{} observations. This enabled us to study its migration. We used spot parameters obtained from LC fitting using \PHOEBE, from different sectors, to quantify the migration (see Table \ref{tab:hotspot}). Using the values of longitudes and the mid-times of the first primary eclipse ($T_\mathrm{seg}$) for each segment, we calculated the rate of change of the longitude. We found that the spot moves $\sim$ $0.696^{\circ}$ per orbital cycle of the inner binary. The spot moves from a longitude of $0^{\circ}-180^{\circ}$ in 2.5 years. Extrapolating this, we get a spot migration period of 5 years. The migration is most probably caused by differential rotation because the spot in question is near the poles, as seen in low $c^\mathrm{spot}$ values (Fig.\ref{fig:spots}).
\citet{petitspotdiffrot} represents the differential rotation as,
\begin{equation}
    \Omega(c^\mathrm{spot}) = \Omega_\mathrm{eq} - \Delta\Omega \cos^{2}(c^\mathrm{spot})
\end{equation}
where, $\Omega$ is the differential rotation as a function of co-latitude (or latitude), $\Omega_\mathrm{eq}$ is the rotation at the equator, and $\Delta\Omega$ is the difference in rotation rate between the pole and
the equator.  Calculating an estimate of $\Delta\Omega$ in BD44Ab gives us $\Delta\Omega = 0.0044 \pm 0.0004 \, \mathrm{rad\,d^{-1}}$. This small differential rotation has been seen in the K1 sub-giant primary in the RS CVn system HR 1099 \citep{petitspotdiffrot} and also K-type main-sequence of V471 Tau \citep{hussainspotdiffrot} with $\Delta\Omega = 0.0152 \pm 0.0008 \, \mathrm{rad\,d^{-1}}$ and $\Delta\Omega = 0.0016 \pm 0.0060\, \mathrm{rad\,d^{-1}}$ respectively. 

\begin{table}
 \caption{Spot parameters for the coldest spot on the secondary of BD44. The parameters have been obtained for each sector using \PHOEBE{} modelling.}
 \label{tab:hotspot}
 \centering
 \begin{tabular}{cccc}
  \hline
   Parameters &  S16 & S22 & S49\\
  \hline

  $T_\mathrm{seg}$ [BJD-2457000] & 1740.8270 & 1900.5753 & 2640.2640  \\
$r^\mathrm{spot}_\mathrm{Ab}$ [deg] &42.846 & 37.311 &39.611\\

 $T^\mathrm{spot}_\mathrm{Ab}$ &0.8988& 0.9296 &0.8500\\

 $c^\mathrm{spot}_\mathrm{Ab}$ [deg] &22.000& 35.745 &21.913\\

  $l^\mathrm{spot}_\mathrm{Ab}$ [deg] &0.000& 29.975  &180.233\\
  
  \hline
 \end{tabular}
 \end{table}

\section{Conclusions}

We obtained independent measurements of different parameters for two triple-lined CHT. Using LC modelling, RV modelling, and \spd{} followed by spectral analysis, we obtained stellar, orbital and atmospheric parameters of all the six stars in the two CHT.  A multi-parameter isochrone fitting constrained the ages of the two systems to be in the order of Gyrs. Isochrones, along with abundances obtained from the disentangled spectra, helped us classify the evolutionary state of the tertiary in the two systems. Furthermore, we gathered the following information about the two systems:\\

\begin{itemize}

    \item \textbf{KIC65:}  The period-ratio of the CHT is $\approx$ 122. All stars in the system are main-sequence stars. The system is a metal-poor one and is $\alpha$-enhanced. The tertiary has 5 possible configurations of mutual inclination with a possibility of near co-planar orbit. Due to the comparatively smaller mass of the tertiary and a wider orbit, the system is stable in the long-term for all values of mutual inclinations. The distance estimated in our study is consistent with the distance obtained in Gaia-DR3. 
    
    \item \textbf{BD44:} This system is a relatively tighter CHT with a period-ratio $\approx$ 73. But still, this system is well above the dynamical instability\footnote{This limit is derived for prograde co-planar motion.} limit as defined in \cite{mardlingaasreth}. The system consists of one main-sequence star (almost at the turn-off) and two sub-giant stars. The abundance patterns of the two sub-giants are similar while the main-sequence differs in Mg and Si abundances. The system has large and cold spots which affect the measurement of some of our parameters. But using the spots we were able to calculate a differential rotation in the sub-giant of the inner binary.  This sub-giant component of the inner pair also contributes significantly to tidal forces. Numerical simulations with tidal interactions show that the inner binary will collide/merge in a few hundred of Myrs due to the radius of the sub-giant exceeding the Roche limit. This leaves behind a wide binary unless the formation of a contact binary stabilises the system. The tertiary does hasten this merger but the effects are drastic if the tertiary is orbiting in an orbit perpendicular to the inner binary orbit.
\end{itemize}

Both the targets can benefit from further photometric and spectroscopic observations which will improve the estimate of obtained parameters. The photometric observations themselves will be quite crucial to check for inclination variations of the inner binary and therefore will give better constraints on the mutual inclination. This will be helpful in constraining evolution scenarios of the CHT as well as star-formation scenarios of these close triples. Nevertheless, this study shows that the use of parameters obtained using independent observations is crucial for the realistic modelling of CHT.

\section*{Acknowledgements}
The authors thank the referee for the invaluable comments and suggestions.
The authors thank Dr. Kyle Conroy and Dr. Andrej Pr\v{s}a for their valuable suggestions and help with \PHOEBE. The authors also acknowledge Dr. Hanno Rein for his help in setting up \textsc{rebound} simulations. This work is funded by the Polish National Science Centre (NCN) through grant 2021/41/N/ST9/02746.
A.M., F.M., T.P., and M.K. are supported by NCN through grant no. 2017/27/B/ST9/02727. F.M. gratefully acknowledges support from the NASA TESS Guest Investigator grant 80NSSC22K0180 (PI A.~Pr\v{s}a).
This paper includes data collected with the TESS mission, obtained from the Mikulski Archive for Space Telescopes (MAST) data archive at the Space Telescope Science Institute (STScI). Funding for the TESS mission is provided by the NASA Explorer Program. STScI is operated by the Association of Universities for Research in Astronomy, Inc., under NASA contract NAS 5–26555. This work also presents results from the European Space Agency (ESA) space mission Gaia. Gaia data are being processed by the Gaia Data Processing and Analysis Consortium (DPAC). Funding for the DPAC is provided by national institutions, in particular, the institutions participating in the Gaia MultiLateral Agreement (MLA). 
\section*{Data Availability}

 The \TESS{} data used in this article are public and hosted on MAST. The data can be accessed from \url{http://dx.doi.org/10.17909/fp8v-5705}. 
 The spectroscopic data will be made available upon request.



\bibliographystyle{mnras}
\bibliography{triplelined} 

\begin{thebibliography}{}
\makeatletter
\relax
\def\mn@urlcharsother{\let\do\@makeother \do\$\do\&\do\#\do\^\do\_\do\%\do\~}
\def\mn@doi{\begingroup\mn@urlcharsother \@ifnextchar [ {\mn@doi@}
  {\mn@doi@[]}}
\def\mn@doi@[#1]#2{\def\@tempa{#1}\ifx\@tempa\@empty \href
  {http://dx.doi.org/#2} {doi:#2}\else \href {http://dx.doi.org/#2} {#1}\fi
  \endgroup}
\def\mn@eprint#1#2{\mn@eprint@#1:#2::\@nil}
\def\mn@eprint@arXiv#1{\href {http://arxiv.org/abs/#1} {{\tt arXiv:#1}}}
\def\mn@eprint@dblp#1{\href {http://dblp.uni-trier.de/rec/bibtex/#1.xml}
  {dblp:#1}}
\def\mn@eprint@#1:#2:#3:#4\@nil{\def\@tempa {#1}\def\@tempb {#2}\def\@tempc
  {#3}\ifx \@tempc \@empty \let \@tempc \@tempb \let \@tempb \@tempa \fi \ifx
  \@tempb \@empty \def\@tempb {arXiv}\fi \@ifundefined
  {mn@eprint@\@tempb}{\@tempb:\@tempc}{\expandafter \expandafter \csname
  mn@eprint@\@tempb\endcsname \expandafter{\@tempc}}}

\bibitem[\protect\citeauthoryear{{Akashi} \& {Soker}}{{Akashi} \&
  {Soker}}{2017}]{pneakashi}
{Akashi} M.,  {Soker} N.,  2017, \mn@doi [\mnras] {10.1093/mnras/stx1058},
  \href {https://ui.adsabs.harvard.edu/abs/2017MNRAS.469.3296A} {469, 3296}

\bibitem[\protect\citeauthoryear{{Asplund}, {Grevesse}, {Sauval}  \&
  {Scott}}{{Asplund} et~al.}{2009}]{2009ARA&A..47..481A}
{Asplund} M.,  {Grevesse} N.,  {Sauval} A.~J.,   {Scott} P.,  2009, \mn@doi
  [\araa] {10.1146/annurev.astro.46.060407.145222}, \href
  {https://ui.adsabs.harvard.edu/abs/2009ARA&A..47..481A} {47, 481}

\bibitem[\protect\citeauthoryear{{Bianchi}, {Herald}, {Efremova}, {Girardi},
  {Zabot}, {Marigo}, {Conti}  \& {Shiao}}{{Bianchi} et~al.}{2011}]{galex}
{Bianchi} L.,  {Herald} J.,  {Efremova} B.,  {Girardi} L.,  {Zabot} A.,
  {Marigo} P.,  {Conti} A.,   {Shiao} B.,  2011, \mn@doi [\apss]
  {10.1007/s10509-010-0581-x}, \href
  {https://ui.adsabs.harvard.edu/abs/2011Ap&SS.335..161B} {335, 161}

\bibitem[\protect\citeauthoryear{{Blanco-Cuaresma}}{{Blanco-Cuaresma}}{2019}]{ispec2019}
{Blanco-Cuaresma} S.,  2019, \mn@doi [\mnras] {10.1093/mnras/stz549}, \href
  {https://ui.adsabs.harvard.edu/abs/2019MNRAS.486.2075B} {486, 2075}

\bibitem[\protect\citeauthoryear{{Blanco-Cuaresma}, {Soubiran}, {Heiter}  \&
  {Jofr{\'e}}}{{Blanco-Cuaresma} et~al.}{2014}]{ispec2014}
{Blanco-Cuaresma} S.,  {Soubiran} C.,  {Heiter} U.,   {Jofr{\'e}} P.,  2014,
  \mn@doi [\aap] {10.1051/0004-6361/201423945}, \href
  {https://ui.adsabs.harvard.edu/abs/2014A&A...569A.111B} {569, A111}

\bibitem[\protect\citeauthoryear{{Borkovits}}{{Borkovits}}{2022}]{borkoCHTreview}
{Borkovits} T.,  2022, \mn@doi [Galaxies] {10.3390/galaxies10010009}, \href
  {https://ui.adsabs.harvard.edu/abs/2022Galax..10....9B} {10, 9}

\bibitem[\protect\citeauthoryear{{Borkovits}, {Rappaport}, {Hajdu}  \&
  {Sztakovics}}{{Borkovits} et~al.}{2015}]{2015borko}
{Borkovits} T.,  {Rappaport} S.,  {Hajdu} T.,   {Sztakovics} J.,  2015, \mn@doi
  [\mnras] {10.1093/mnras/stv015}, \href
  {https://ui.adsabs.harvard.edu/abs/2015MNRAS.448..946B} {448, 946}

\bibitem[\protect\citeauthoryear{{Borkovits}, {Hajdu}, {Sztakovics},
  {Rappaport}, {Levine}, {B{\'\i}r{\'o}}  \& {Klagyivik}}{{Borkovits}
  et~al.}{2016}]{borkovitskeptrip}
{Borkovits} T.,  {Hajdu} T.,  {Sztakovics} J.,  {Rappaport} S.,  {Levine} A.,
  {B{\'\i}r{\'o}} I.~B.,   {Klagyivik} P.,  2016, \mn@doi [\mnras]
  {10.1093/mnras/stv2530}, \href
  {https://ui.adsabs.harvard.edu/abs/2016MNRAS.455.4136B} {455, 4136}

\bibitem[\protect\citeauthoryear{{Borkovits}, {Rappaport}, {Hajdu}, {Maxted},
  {P{\'a}l}, {Forg{\'a}cs-Dajka}, {Klagyivik}  \& {Mitnyan}}{{Borkovits}
  et~al.}{2020}]{2020borko2tt}
{Borkovits} T.,  {Rappaport} S.~A.,  {Hajdu} T.,  {Maxted} P.~F.~L.,  {P{\'a}l}
  A.,  {Forg{\'a}cs-Dajka} E.,  {Klagyivik} P.,   {Mitnyan} T.,  2020, \mn@doi
  [\mnras] {10.1093/mnras/staa495}, \href
  {https://ui.adsabs.harvard.edu/abs/2020MNRAS.493.5005B} {493, 5005}

\bibitem[\protect\citeauthoryear{Borkovits, Rappaport, Toonen, Moe, Mitnyan  \&
  Csányi}{Borkovits et~al.}{2022}]{borkozlk}
Borkovits T.,  Rappaport S.~A.,  Toonen S.,  Moe M.,  Mitnyan T.,   Csányi I.,
   2022, \mn@doi [Monthly Notices of the Royal Astronomical Society]
  {10.1093/mnras/stac1983}

\bibitem[\protect\citeauthoryear{{Box} \& {Jenkins}}{{Box} \&
  {Jenkins}}{1976}]{autocorrtests}
{Box} G. E.~P.,  {Jenkins} G.~M.,  1976, in Holden-Day Series in Time Series
  Analysis.

\bibitem[\protect\citeauthoryear{{Cardelli}, {Clayton}  \& {Mathis}}{{Cardelli}
  et~al.}{1989}]{car89}
{Cardelli} J.~A.,  {Clayton} G.~C.,   {Mathis} J.~S.,  1989, \mn@doi [\apj]
  {10.1086/167900}, \href
  {https://ui.adsabs.harvard.edu/abs/1989ApJ...345..245C} {345, 245}

\bibitem[\protect\citeauthoryear{{Choi}, {Dotter}, {Conroy}, {Cantiello},
  {Paxton}  \& {Johnson}}{{Choi} et~al.}{2016}]{mistc2016}
{Choi} J.,  {Dotter} A.,  {Conroy} C.,  {Cantiello} M.,  {Paxton} B.,
  {Johnson} B.~D.,  2016, \mn@doi [\apj] {10.3847/0004-637X/823/2/102}, \href
  {https://ui.adsabs.harvard.edu/abs/2016ApJ...823..102C} {823, 102}

\bibitem[\protect\citeauthoryear{{Claret} \& {Bloemen}}{{Claret} \&
  {Bloemen}}{2011}]{claretlimbdarkening}
{Claret} A.,  {Bloemen} S.,  2011, \mn@doi [\aap]
  {10.1051/0004-6361/201116451}, \href
  {https://ui.adsabs.harvard.edu/abs/2011A&A...529A..75C} {529, A75}

\bibitem[\protect\citeauthoryear{{Conroy} et~al.,}{{Conroy}
  et~al.}{2020}]{conroy2020}
{Conroy} K.~E.,  et~al., 2020, \mn@doi [\apjs] {10.3847/1538-4365/abb4e2},
  \href {https://ui.adsabs.harvard.edu/abs/2020ApJS..250...34C} {250, 34}

\bibitem[\protect\citeauthoryear{{Dotter}}{{Dotter}}{2016}]{mistd2016}
{Dotter} A.,  2016, \mn@doi [\apjs] {10.3847/0067-0049/222/1/8}, \href
  {https://ui.adsabs.harvard.edu/abs/2016ApJS..222....8D} {222, 8}

\bibitem[\protect\citeauthoryear{{Duch{\^e}ne} \& {Kraus}}{{Duch{\^e}ne} \&
  {Kraus}}{2013}]{multiplicity}
{Duch{\^e}ne} G.,  {Kraus} A.,  2013, \mn@doi [\araa]
  {10.1146/annurev-astro-081710-102602}, \href
  {https://ui.adsabs.harvard.edu/abs/2013ARA&A..51..269D} {51, 269}

\bibitem[\protect\citeauthoryear{{Eggleton}}{{Eggleton}}{1983}]{eggletonrocheradii}
{Eggleton} P.~P.,  1983, \mn@doi [\apj] {10.1086/160960}, \href
  {https://ui.adsabs.harvard.edu/abs/1983ApJ...268..368E} {268, 368}

\bibitem[\protect\citeauthoryear{{Eggleton} \& {Kiseleva-Eggleton}}{{Eggleton}
  \& {Kiseleva-Eggleton}}{2001}]{peternlud2001}
{Eggleton} P.~P.,  {Kiseleva-Eggleton} L.,  2001, \mn@doi [\apj]
  {10.1086/323843}, \href
  {https://ui.adsabs.harvard.edu/abs/2001ApJ...562.1012E} {562, 1012}

\bibitem[\protect\citeauthoryear{{Eggleton} \& {Verbunt}}{{Eggleton} \&
  {Verbunt}}{1986}]{eggletonlmxb}
{Eggleton} P.~P.,  {Verbunt} F.,  1986, \mn@doi [\mnras]
  {10.1093/mnras/220.1.13P}, \href
  {https://ui.adsabs.harvard.edu/abs/1986MNRAS.220P..13E} {220, 13P}

\bibitem[\protect\citeauthoryear{{Eisner} et~al.,}{{Eisner}
  et~al.}{2022}]{eisnercht}
{Eisner} N.~L.,  et~al., 2022, \mn@doi [\mnras] {10.1093/mnras/stab3619}, \href
  {https://ui.adsabs.harvard.edu/abs/2022MNRAS.511.4710E} {511, 4710}

\bibitem[\protect\citeauthoryear{{Ford}, {Kozinsky}  \& {Rasio}}{{Ford}
  et~al.}{2000}]{fordseculartrip}
{Ford} E.~B.,  {Kozinsky} B.,   {Rasio} F.~A.,  2000, \mn@doi [\apj]
  {10.1086/308815}, \href
  {https://ui.adsabs.harvard.edu/abs/2000ApJ...535..385F} {535, 385}

\bibitem[\protect\citeauthoryear{{Foreman-Mackey}, {Hogg}, {Lang}  \&
  {Goodman}}{{Foreman-Mackey} et~al.}{2013}]{emcee2013}
{Foreman-Mackey} D.,  {Hogg} D.~W.,  {Lang} D.,   {Goodman} J.,  2013, \mn@doi
  [\pasp] {10.1086/670067}, \href
  {https://ui.adsabs.harvard.edu/abs/2013PASP..125..306F} {125, 306}

\bibitem[\protect\citeauthoryear{{Foreman-Mackey} et~al.,}{{Foreman-Mackey}
  et~al.}{2019}]{emcee2019}
{Foreman-Mackey} D.,  et~al., 2019, \mn@doi [The Journal of Open Source
  Software] {10.21105/joss.01864}, \href
  {https://ui.adsabs.harvard.edu/abs/2019JOSS....4.1864F} {4, 1864}

\bibitem[\protect\citeauthoryear{Francq \& Zakoïan}{Francq \&
  Zakoïan}{2009}]{bartlett}
Francq C.,  Zakoïan J.-M.,  2009, \mn@doi [Journal of Time Series Analysis]
  {https://doi.org/10.1111/j.1467-9892.2009.00623.x}, 30, 449

\bibitem[\protect\citeauthoryear{{Fuller}, {Derekas}, {Borkovits}, {Huber},
  {Bedding}  \& {Kiss}}{{Fuller} et~al.}{2013}]{fullertidalpulstriples}
{Fuller} J.,  {Derekas} A.,  {Borkovits} T.,  {Huber} D.,  {Bedding} T.~R.,
  {Kiss} L.~L.,  2013, \mn@doi [\mnras] {10.1093/mnras/sts511}, \href
  {https://ui.adsabs.harvard.edu/abs/2013MNRAS.429.2425F} {429, 2425}

\bibitem[\protect\citeauthoryear{{Gaia Collaboration} et~al.,}{{Gaia
  Collaboration} et~al.}{2022}]{gaia2022}
{Gaia Collaboration} et~al., 2022, arXiv e-prints, \href
  {https://ui.adsabs.harvard.edu/abs/2022arXiv220800211G} {p. arXiv:2208.00211}

\bibitem[\protect\citeauthoryear{{Gilmore} et~al.,}{{Gilmore}
  et~al.}{2012}]{gso2012}
{Gilmore} G.,  et~al., 2012, The Messenger, \href
  {https://ui.adsabs.harvard.edu/abs/2012Msngr.147...25G} {147, 25}

\bibitem[\protect\citeauthoryear{{Gray}}{{Gray}}{2005}]{stellarphotgray}
{Gray} D.~F.,  2005, {The Observation and Analysis of Stellar Photospheres}.
Cambridge University Press

\bibitem[\protect\citeauthoryear{{Grevesse}, {Asplund}  \& {Sauval}}{{Grevesse}
  et~al.}{2007}]{grevesse2007}
{Grevesse} N.,  {Asplund} M.,   {Sauval} A.~J.,  2007, \mn@doi [\ssr]
  {10.1007/s11214-007-9173-7}, \href
  {https://ui.adsabs.harvard.edu/abs/2007SSRv..130..105G} {130, 105}

\bibitem[\protect\citeauthoryear{Gronchi \& Tommei}{Gronchi \&
  Tommei}{2007}]{mut_i_from_i1_i2}
Gronchi G.~F.,  Tommei G.,  2007, Discrete and Continuous Dynamical Systems -
  B, 7, 755

\bibitem[\protect\citeauthoryear{{Gustafsson}, {Edvardsson}, {Eriksson},
  {J{\o}rgensen}, {Nordlund}  \& {Plez}}{{Gustafsson} et~al.}{2008}]{marcges}
{Gustafsson} B.,  {Edvardsson} B.,  {Eriksson} K.,  {J{\o}rgensen} U.~G.,
  {Nordlund} {\r{A}}.,   {Plez} B.,  2008, \mn@doi [\aap]
  {10.1051/0004-6361:200809724}, \href
  {https://ui.adsabs.harvard.edu/abs/2008A&A...486..951G} {486, 951}

\bibitem[\protect\citeauthoryear{{Hadrava}}{{Hadrava}}{1995}]{hadrava_disent}
{Hadrava} P.,  1995, \aaps, \href
  {https://ui.adsabs.harvard.edu/abs/1995A&AS..114..393H} {114, 393}

\bibitem[\protect\citeauthoryear{{Hamers} \& {Dosopoulou}}{{Hamers} \&
  {Dosopoulou}}{2019}]{masstranslk}
{Hamers} A.~S.,  {Dosopoulou} F.,  2019, \mn@doi [\apj]
  {10.3847/1538-4357/ab001d}, \href
  {https://ui.adsabs.harvard.edu/abs/2019ApJ...872..119H} {872, 119}

\bibitem[\protect\citeauthoryear{{He{\l}miniak}, {Ukita}, {Kambe},
  {Koz{\l}owski}, {Sybilski}, {Ratajczak}, {Maehara}  \&
  {Konacki}}{{He{\l}miniak} et~al.}{2016}]{krishides2016}
{He{\l}miniak} K.~G.,  {Ukita} N.,  {Kambe} E.,  {Koz{\l}owski} S.~K.,
  {Sybilski} P.,  {Ratajczak} M.,  {Maehara} H.,   {Konacki} M.,  2016, \mn@doi
  [\mnras] {10.1093/mnras/stw1514}, \href
  {https://ui.adsabs.harvard.edu/abs/2016MNRAS.461.2896H} {461, 2896}

\bibitem[\protect\citeauthoryear{{He{\l}miniak} et~al.,}{{He{\l}miniak}
  et~al.}{2017}]{krishides2017}
{He{\l}miniak} K.~G.,  et~al., 2017, \mn@doi [\mnras] {10.1093/mnras/stx385},
  \href {https://ui.adsabs.harvard.edu/abs/2017MNRAS.468.1726H} {468, 1726}

\bibitem[\protect\citeauthoryear{{He{\l}miniak} et~al.,}{{He{\l}miniak}
  et~al.}{2021}]{hel21}
{He{\l}miniak} K.~G.,  et~al., 2021, \mn@doi [\mnras] {10.1093/mnras/stab2963},
  \href {https://ui.adsabs.harvard.edu/abs/2021MNRAS.508.5687H} {508, 5687}

\bibitem[\protect\citeauthoryear{{Holmgren}, {Hadrava}, {Harmanec}, {Eenens},
  {Corral}, {Yang}, {Ak}  \& {Bozi{\'c}}}{{Holmgren}
  et~al.}{1999}]{holmgrenlowlightfrac}
{Holmgren} D.~E.,  {Hadrava} P.,  {Harmanec} P.,  {Eenens} P.,  {Corral} L.~J.,
   {Yang} S.,  {Ak} H.,   {Bozi{\'c}} H.,  1999, \aap, \href
  {https://ui.adsabs.harvard.edu/abs/1999A&A...345..855H} {345, 855}

\bibitem[\protect\citeauthoryear{{Horvat}, {Conroy}, {Pablo}, {Hambleton},
  {Kochoska}, {Giammarco}  \& {Pr{\v{s}}a}}{{Horvat} et~al.}{2018}]{phoebe2018}
{Horvat} M.,  {Conroy} K.~E.,  {Pablo} H.,  {Hambleton} K.~M.,  {Kochoska} A.,
  {Giammarco} J.,   {Pr{\v{s}}a} A.,  2018, \mn@doi [\apjs]
  {10.3847/1538-4365/aacd0f}, \href
  {https://ui.adsabs.harvard.edu/abs/2018ApJS..237...26H} {237, 26}

\bibitem[\protect\citeauthoryear{{Howell} et~al.,}{{Howell}
  et~al.}{2014}]{k2mission}
{Howell} S.~B.,  et~al., 2014, \mn@doi [\pasp] {10.1086/676406}, \href
  {https://ui.adsabs.harvard.edu/abs/2014PASP..126..398H} {126, 398}

\bibitem[\protect\citeauthoryear{{Hussain}, {Allende Prieto}, {Saar}  \&
  {Still}}{{Hussain} et~al.}{2006}]{hussainspotdiffrot}
{Hussain} G.~A.~J.,  {Allende Prieto} C.,  {Saar} S.~H.,   {Still} M.,  2006,
  \mn@doi [\mnras] {10.1111/j.1365-2966.2006.10073.x}, \href
  {https://ui.adsabs.harvard.edu/abs/2006MNRAS.367.1699H} {367, 1699}

\bibitem[\protect\citeauthoryear{{Hut}}{{Hut}}{1981}]{hut1981}
{Hut} P.,  1981, \aap, \href
  {https://ui.adsabs.harvard.edu/abs/1981A&A....99..126H} {99, 126}

\bibitem[\protect\citeauthoryear{{Ilijic}, {Hensberge}, {Pavlovski}  \&
  {Freyhammer}}{{Ilijic} et~al.}{2004}]{fdbinary}
{Ilijic} S.,  {Hensberge} H.,  {Pavlovski} K.,   {Freyhammer} L.~M.,  2004, in
  {Hilditch} R.~W.,  {Hensberge} H.,   {Pavlovski} K.,  eds,  Astronomical
  Society of the Pacific Conference Series Vol. 318, Spectroscopically and
  Spatially Resolving the Components of the Close Binary Stars. pp 111--113

\bibitem[\protect\citeauthoryear{{Izumiura}}{{Izumiura}}{1999}]{izumiurahides}
{Izumiura} H.,  1999, in {Chen} P.~S.,  ed., Proc. 4th East Asian Meeting on
  Astronomy, Observational Astrophysics in Asia and its Future. Kunming Yunnan
  Observatory, p.~77

\bibitem[\protect\citeauthoryear{Jones, Pejcha  \& Corradi}{Jones
  et~al.}{2019}]{pnebfromtrip}
Jones D.,  Pejcha O.,   Corradi R. L.~M.,  2019, \mn@doi [\mnras]
  {10.1093/mnras/stz2293}, 489, 2195

\bibitem[\protect\citeauthoryear{{Jones} et~al.,}{{Jones}
  et~al.}{2020}]{phoebeJ2020}
{Jones} D.,  et~al., 2020, \mn@doi [\apjs] {10.3847/1538-4365/ab7927}, \href
  {https://ui.adsabs.harvard.edu/abs/2020ApJS..247...63J} {247, 63}

\bibitem[\protect\citeauthoryear{{Kambe} et~al.,}{{Kambe}
  et~al.}{2013}]{kambe2013}
{Kambe} E.,  et~al., 2013, \mn@doi [\pasj] {10.1093/pasj/65.1.15}, \href
  {https://ui.adsabs.harvard.edu/abs/2013PASJ...65...15K} {65, 15}

\bibitem[\protect\citeauthoryear{{Kervella}, {Th{\'e}venin}, {Di Folco}  \&
  {S{\'e}gransan}}{{Kervella} et~al.}{2004}]{ker04}
{Kervella} P.,  {Th{\'e}venin} F.,  {Di Folco} E.,   {S{\'e}gransan} D.,  2004,
  \mn@doi [\aap] {10.1051/0004-6361:20035930}, \href
  {https://ui.adsabs.harvard.edu/abs/2004A&A...426..297K} {426, 297}

\bibitem[\protect\citeauthoryear{{Kobulnicky} \& {Fryer}}{{Kobulnicky} \&
  {Fryer}}{2007}]{kobulnicky2007}
{Kobulnicky} H.~A.,  {Fryer} C.~L.,  2007, \mn@doi [\apj] {10.1086/522073},
  \href {https://ui.adsabs.harvard.edu/abs/2007ApJ...670..747K} {670, 747}

\bibitem[\protect\citeauthoryear{{Konacki}, {Muterspaugh}, {Kulkarni}  \&
  {He{\l}miniak}}{{Konacki} et~al.}{2010}]{v2fit}
{Konacki} M.,  {Muterspaugh} M.~W.,  {Kulkarni} S.~R.,   {He{\l}miniak} K.~G.,
  2010, \mn@doi [\apj] {10.1088/0004-637X/719/2/1293}, \href
  {https://ui.adsabs.harvard.edu/abs/2010ApJ...719.1293K} {719, 1293}

\bibitem[\protect\citeauthoryear{{Korth}, {Moharana}, {Pe{\v{s}}ta},
  {Czavalinga}  \& {Conroy}}{{Korth} et~al.}{2021}]{korthmoh}
{Korth} J.,  {Moharana} A.,  {Pe{\v{s}}ta} M.,  {Czavalinga} D.~R.,   {Conroy}
  K.~E.,  2021, \mn@doi [Contributions of the Astronomical Observatory Skalnate
  Pleso] {10.31577/caosp.2021.51.1.58}, \href
  {https://ui.adsabs.harvard.edu/abs/2021CoSka..51...58K} {51, 58}

\bibitem[\protect\citeauthoryear{{Kozai}}{{Kozai}}{1962}]{kozai}
{Kozai} Y.,  1962, \mn@doi [\aj] {10.1086/108790}, \href
  {https://ui.adsabs.harvard.edu/abs/1962AJ.....67..591K} {67, 591}

\bibitem[\protect\citeauthoryear{{Lee}, {Offner}, {Kratter}, {Smullen}  \&
  {Li}}{{Lee} et~al.}{2019}]{Leemultipleform}
{Lee} A.~T.,  {Offner} S. S.~R.,  {Kratter} K.~M.,  {Smullen} R.~A.,   {Li}
  P.~S.,  2019, \mn@doi [\apj] {10.3847/1538-4357/ab584b}, \href
  {https://ui.adsabs.harvard.edu/abs/2019ApJ...887..232L} {887, 232}

\bibitem[\protect\citeauthoryear{{Lidov}}{{Lidov}}{1962}]{lidov}
{Lidov} M.~L.,  1962, \mn@doi [\planss] {10.1016/0032-0633(62)90129-0}, \href
  {https://ui.adsabs.harvard.edu/abs/1962P&SS....9..719L} {9, 719}

\bibitem[\protect\citeauthoryear{{Maoz}, {Mannucci}  \& {Nelemans}}{{Maoz}
  et~al.}{2014}]{maozwdsne}
{Maoz} D.,  {Mannucci} F.,   {Nelemans} G.,  2014, \mn@doi [\araa]
  {10.1146/annurev-astro-082812-141031}, \href
  {https://ui.adsabs.harvard.edu/abs/2014ARA&A..52..107M} {52, 107}

\bibitem[\protect\citeauthoryear{{Marcadon}, {He{\l}miniak}, {Marques},
  {Paw{\l}aszek}, {Sybilski}, {Koz{\l}owski}, {Ratajczak}  \&
  {Konacki}}{{Marcadon} et~al.}{2020}]{fredv12}
{Marcadon} F.,  {He{\l}miniak} K.~G.,  {Marques} J.~P.,  {Paw{\l}aszek} R.,
  {Sybilski} P.,  {Koz{\l}owski} S.~K.,  {Ratajczak} M.,   {Konacki} M.,  2020,
  \mn@doi [\mnras] {10.1093/mnras/staa3040}, \href
  {https://ui.adsabs.harvard.edu/abs/2020MNRAS.499.3019M} {499, 3019}

\bibitem[\protect\citeauthoryear{{Mardling} \& {Aarseth}}{{Mardling} \&
  {Aarseth}}{2001}]{mardlingaasreth}
{Mardling} R.~A.,  {Aarseth} S.~J.,  2001, \mn@doi [\mnras]
  {10.1046/j.1365-8711.2001.03974.x}, \href
  {https://ui.adsabs.harvard.edu/abs/2001MNRAS.321..398M} {321, 398}

\bibitem[\protect\citeauthoryear{{Mason}, {Hartkopf}, {Gies}, {Henry}  \&
  {Helsel}}{{Mason} et~al.}{2009}]{Mason2009}
{Mason} B.~D.,  {Hartkopf} W.~I.,  {Gies} D.~R.,  {Henry} T.~J.,   {Helsel}
  J.~W.,  2009, \mn@doi [\aj] {10.1088/0004-6256/137/2/3358}, \href
  {https://ui.adsabs.harvard.edu/abs/2009AJ....137.3358M} {137, 3358}

\bibitem[\protect\citeauthoryear{{Maxted} et~al.,}{{Maxted}
  et~al.}{2020}]{maxted2020}
{Maxted} P.~F.~L.,  et~al., 2020, \mn@doi [\mnras] {10.1093/mnras/staa1662},
  \href {https://ui.adsabs.harvard.edu/abs/2020MNRAS.498..332M} {498, 332}

\bibitem[\protect\citeauthoryear{{Mayer}, {Harmanec}  \& {Pavlovski}}{{Mayer}
  et~al.}{2013}]{2013mayerlowlightfrac}
{Mayer} P.,  {Harmanec} P.,   {Pavlovski} K.,  2013, \mn@doi [\aap]
  {10.1051/0004-6361/201220388}, \href
  {https://ui.adsabs.harvard.edu/abs/2013A&A...550A...2M} {550, A2}

\bibitem[\protect\citeauthoryear{{Moe} \& {Kratter}}{{Moe} \&
  {Kratter}}{2018}]{maxwellkratter}
{Moe} M.,  {Kratter} K.~M.,  2018, \mn@doi [\apj] {10.3847/1538-4357/aaa6d2},
  \href {https://ui.adsabs.harvard.edu/abs/2018ApJ...854...44M} {854, 44}

\bibitem[\protect\citeauthoryear{{Naoz} \& {Fabrycky}}{{Naoz} \&
  {Fabrycky}}{2014}]{naozobliqtrip}
{Naoz} S.,  {Fabrycky} D.~C.,  2014, \mn@doi [\apj]
  {10.1088/0004-637X/793/2/137}, \href
  {https://ui.adsabs.harvard.edu/abs/2014ApJ...793..137N} {793, 137}

\bibitem[\protect\citeauthoryear{Nelder \& Mead}{Nelder \&
  Mead}{1965}]{neldermead}
Nelder J.~A.,  Mead R.,  1965, \mn@doi [The Computer Journal]
  {10.1093/comjnl/7.4.308}, 7, 308

\bibitem[\protect\citeauthoryear{{Paxton}, {Bildsten}, {Dotter}, {Herwig},
  {Lesaffre}  \& {Timmes}}{{Paxton} et~al.}{2011}]{mesa2011}
{Paxton} B.,  {Bildsten} L.,  {Dotter} A.,  {Herwig} F.,  {Lesaffre} P.,
  {Timmes} F.,  2011, \mn@doi [\apjs] {10.1088/0067-0049/192/1/3}, \href
  {https://ui.adsabs.harvard.edu/abs/2011ApJS..192....3P} {192, 3}

\bibitem[\protect\citeauthoryear{{Paxton} et~al.,}{{Paxton}
  et~al.}{2013}]{mesa2013}
{Paxton} B.,  et~al., 2013, \mn@doi [\apjs] {10.1088/0067-0049/208/1/4}, \href
  {https://ui.adsabs.harvard.edu/abs/2013ApJS..208....4P} {208, 4}

\bibitem[\protect\citeauthoryear{{Paxton} et~al.,}{{Paxton}
  et~al.}{2015}]{mesa2015}
{Paxton} B.,  et~al., 2015, \mn@doi [\apjs] {10.1088/0067-0049/220/1/15}, \href
  {https://ui.adsabs.harvard.edu/abs/2015ApJS..220...15P} {220, 15}

\bibitem[\protect\citeauthoryear{{Paxton} et~al.,}{{Paxton}
  et~al.}{2018}]{mesa2018}
{Paxton} B.,  et~al., 2018, \mn@doi [\apjs] {10.3847/1538-4365/aaa5a8}, \href
  {https://ui.adsabs.harvard.edu/abs/2018ApJS..234...34P} {234, 34}

\bibitem[\protect\citeauthoryear{{Perets} \& {Fabrycky}}{{Perets} \&
  {Fabrycky}}{2009}]{peretsfab}
{Perets} H.~B.,  {Fabrycky} D.~C.,  2009, \mn@doi [\apj]
  {10.1088/0004-637X/697/2/1048}, \href
  {https://ui.adsabs.harvard.edu/abs/2009ApJ...697.1048P} {697, 1048}

\bibitem[\protect\citeauthoryear{{Petit} et~al.,}{{Petit}
  et~al.}{2004}]{petitspotdiffrot}
{Petit} P.,  et~al., 2004, \mn@doi [\mnras] {10.1111/j.1365-2966.2004.07420.x},
  \href {https://ui.adsabs.harvard.edu/abs/2004MNRAS.348.1175P} {348, 1175}

\bibitem[\protect\citeauthoryear{{Pr{\v{s}}a} et~al.,}{{Pr{\v{s}}a}
  et~al.}{2016}]{phoebe2016}
{Pr{\v{s}}a} A.,  et~al., 2016, \mn@doi [\apjs] {10.3847/1538-4365/227/2/29},
  \href {https://ui.adsabs.harvard.edu/abs/2016ApJS..227...29P} {227, 29}

\bibitem[\protect\citeauthoryear{{Raghavan} et~al.,}{{Raghavan}
  et~al.}{2010}]{solarmultiplesurv}
{Raghavan} D.,  et~al., 2010, \mn@doi [\apjs] {10.1088/0067-0049/190/1/1},
  \href {https://ui.adsabs.harvard.edu/abs/2010ApJS..190....1R} {190, 1}

\bibitem[\protect\citeauthoryear{{Randich}, {Gilmore}  \& {Gaia-ESO
  Consortium}}{{Randich} et~al.}{2013}]{gso2013}
{Randich} S.,  {Gilmore} G.,   {Gaia-ESO Consortium} 2013, The Messenger, \href
  {https://ui.adsabs.harvard.edu/abs/2013Msngr.154...47R} {154, 47}

\bibitem[\protect\citeauthoryear{{Rappaport}, {Deck}, {Levine}, {Borkovits},
  {Carter}, {El Mellah}, {Sanchis-Ojeda}  \& {Kalomeni}}{{Rappaport}
  et~al.}{2013}]{rappa2013}
{Rappaport} S.,  {Deck} K.,  {Levine} A.,  {Borkovits} T.,  {Carter} J.,  {El
  Mellah} I.,  {Sanchis-Ojeda} R.,   {Kalomeni} B.,  2013, \mn@doi [\apj]
  {10.1088/0004-637X/768/1/33}, \href
  {https://ui.adsabs.harvard.edu/abs/2013ApJ...768...33R} {768, 33}

\bibitem[\protect\citeauthoryear{{Rappaport} et~al.,}{{Rappaport}
  et~al.}{2022}]{rappa6cht2022}
{Rappaport} S.~A.,  et~al., 2022, \mn@doi [\mnras] {10.1093/mnras/stac957},
  \href {https://ui.adsabs.harvard.edu/abs/2022MNRAS.513.4341R} {513, 4341}

\bibitem[\protect\citeauthoryear{{Rein} \& {Liu}}{{Rein} \&
  {Liu}}{2012}]{rebound}
{Rein} H.,  {Liu} S.~F.,  2012, \mn@doi [\aap] {10.1051/0004-6361/201118085},
  \href {https://ui.adsabs.harvard.edu/abs/2012A&A...537A.128R} {537, A128}

\bibitem[\protect\citeauthoryear{{Rein} \& {Tamayo}}{{Rein} \&
  {Tamayo}}{2015}]{whfast}
{Rein} H.,  {Tamayo} D.,  2015, \mn@doi [\mnras] {10.1093/mnras/stv1257}, \href
  {https://ui.adsabs.harvard.edu/abs/2015MNRAS.452..376R} {452, 376}

\bibitem[\protect\citeauthoryear{{Ricker} et~al.,}{{Ricker}
  et~al.}{2015}]{TESSricker}
{Ricker} G.~R.,  et~al., 2015, \mn@doi [Journal of Astronomical Telescopes,
  Instruments, and Systems] {10.1117/1.JATIS.1.1.014003}, \href
  {https://ui.adsabs.harvard.edu/abs/2015JATIS...1a4003R} {1, 014003}

\bibitem[\protect\citeauthoryear{{Rucinski}}{{Rucinski}}{1999a}]{bf_discussion}
{Rucinski} S.,  1999a, Turkish Journal of Physics, \href
  {https://ui.adsabs.harvard.edu/abs/1999TJPh...23..271R} {23, 271}

\bibitem[\protect\citeauthoryear{{Rucinski}}{{Rucinski}}{1999b}]{bfsvd}
{Rucinski} S.,  1999b, in {Hearnshaw} J.~B.,  {Scarfe} C.~D.,  eds,
  Astronomical Society of the Pacific Conference Series Vol. 185, IAU Colloq.
  170: Precise Stellar Radial Velocities. p.~82 (\mn@eprint {arXiv}
  {astro-ph/9807327})

\bibitem[\protect\citeauthoryear{{Shatsky} \& {Tokovinin}}{{Shatsky} \&
  {Tokovinin}}{2002}]{shatskyntoko}
{Shatsky} N.,  {Tokovinin} A.,  2002, \mn@doi [\aap]
  {10.1051/0004-6361:20011542}, \href
  {https://ui.adsabs.harvard.edu/abs/2002A&A...382...92S} {382, 92}

\bibitem[\protect\citeauthoryear{{Simon} \& {Sturm}}{{Simon} \&
  {Sturm}}{1994}]{simonsttrum_disentangling}
{Simon} K.~P.,  {Sturm} E.,  1994, \aap, \href
  {https://ui.adsabs.harvard.edu/abs/1994A&A...281..286S} {281, 286}

\bibitem[\protect\citeauthoryear{{Tamayo}, {Rein}, {Shi}  \&
  {Hernandez}}{{Tamayo} et~al.}{2020}]{reboundx}
{Tamayo} D.,  {Rein} H.,  {Shi} P.,   {Hernandez} D.~M.,  2020, \mn@doi
  [\mnras] {10.1093/mnras/stz2870}, \href
  {https://ui.adsabs.harvard.edu/abs/2020MNRAS.491.2885T} {491, 2885}

\bibitem[\protect\citeauthoryear{{Tokovinin}}{{Tokovinin}}{2004}]{tokostats2004}
{Tokovinin} A.,  2004, in {Allen} C.,  {Scarfe} C.,  eds,  Revista Mexicana de
  Astronomia y Astrofisica Conference Series Vol. 21, Revista Mexicana de
  Astronomia y Astrofisica Conference Series. pp 7--14

\bibitem[\protect\citeauthoryear{{Tokovinin}}{{Tokovinin}}{2017}]{tokoorbittriples}
{Tokovinin} A.,  2017, \mn@doi [\apj] {10.3847/1538-4357/aa7746}, \href
  {https://ui.adsabs.harvard.edu/abs/2017ApJ...844..103T} {844, 103}

\bibitem[\protect\citeauthoryear{{Tokovinin}}{{Tokovinin}}{2021}]{tokoformationreview}
{Tokovinin} A.,  2021, \mn@doi [Universe] {10.3390/universe7090352}, \href
  {https://ui.adsabs.harvard.edu/abs/2021Univ....7..352T} {7, 352}

\bibitem[\protect\citeauthoryear{{Tokovinin}}{{Tokovinin}}{2022}]{tokogaiatrip}
{Tokovinin} A.,  2022, \mn@doi [\apj] {10.3847/1538-4357/ac4584}, \href
  {https://ui.adsabs.harvard.edu/abs/2022ApJ...926....1T} {926, 1}

\bibitem[\protect\citeauthoryear{{Toonen}, {Portegies Zwart}, {Hamers}  \&
  {Bandopadhyay}}{{Toonen} et~al.}{2020}]{toonen2020}
{Toonen} S.,  {Portegies Zwart} S.,  {Hamers} A.~S.,   {Bandopadhyay} D.,
  2020, \mn@doi [\aap] {10.1051/0004-6361/201936835}, \href
  {https://ui.adsabs.harvard.edu/abs/2020A&A...640A..16T} {640, A16}

\bibitem[\protect\citeauthoryear{{Torres}, {Andersen}  \&
  {Gim{\'e}nez}}{{Torres} et~al.}{2010}]{torresaccuracy}
{Torres} G.,  {Andersen} J.,   {Gim{\'e}nez} A.,  2010, \mn@doi [\aapr]
  {10.1007/s00159-009-0025-1}, \href
  {https://ui.adsabs.harvard.edu/abs/2010A&ARv..18...67T} {18, 67}

\bibitem[\protect\citeauthoryear{{Voges} et~al.,}{{Voges} et~al.}{1999}]{rosat}
{Voges} W.,  et~al., 1999, \aap, \href
  {https://ui.adsabs.harvard.edu/abs/1999A&A...349..389V} {349, 389}

\bibitem[\protect\citeauthoryear{{Zasche} \& {Paschke}}{{Zasche} \&
  {Paschke}}{2012}]{hyhya}
{Zasche} P.,  {Paschke} A.,  2012, \mn@doi [\aap]
  {10.1051/0004-6361/201219392}, \href
  {https://ui.adsabs.harvard.edu/abs/2012A&A...542L..23Z} {542, L23}

\bibitem[\protect\citeauthoryear{Zasche, Vokrouhlický, Barlow  \&
  Mašek}{Zasche et~al.}{2023}]{Zasche_2023}
Zasche P.,  Vokrouhlický D.,  Barlow B.~N.,   Mašek M.,  2023, \mn@doi [The
  Astronomical Journal] {10.3847/1538-3881/acae95}, 165, 81

\bibitem[\protect\citeauthoryear{{Zucker} \& {Mazeh}}{{Zucker} \&
  {Mazeh}}{1994}]{zuckermazeh}
{Zucker} S.,  {Mazeh} T.,  1994, \mn@doi [\apj] {10.1086/173605}, \href
  {https://ui.adsabs.harvard.edu/abs/1994ApJ...420..806Z} {420, 806}

\bibitem[\protect\citeauthoryear{{von Zeipel}}{{von Zeipel}}{1910}]{zeipel}
{von Zeipel} H.,  1910, \mn@doi [Astronomische Nachrichten]
  {10.1002/asna.19091832202}, \href
  {https://ui.adsabs.harvard.edu/abs/1910AN....183..345V} {183, 345}

\makeatother
\end{thebibliography}




\appendix

\section{Distance and reddening estimates from the isochrones}
 \label{appx:isochrones}
The reddening-free distance $d_0$ are estimated simultaneously with the reddening $E(B-V)$, using the available observed total magnitudes in different filters, and the predicted total brightness of the system in the same filters, for a given triplet of points (= stellar masses) on the same isochrone. We use the $T_{\rm_{eff}}$--surface brightness relations from \citet{ker04}, to calculate the distances $d_\lambda$ and distance moduli $(m-M)_\lambda = 5 \log(d_\lambda) + 5$ in each band. To obtain the extinction-free modulus $(m-M)_0$ we fit a straight line on the $A_\lambda$ vs. $(m-M)_\lambda$ plane, where the $A_\lambda$ are extinction coefficients in each band. We followed the extinction law of \citet{car89} $A_U:A_B:A_V:A_R:A_I:A_J:A_H:A_K = 4.855:4.064:3.1:2.545:1.801:0.88:0.558:0.36$, which assumes $R_V=3.1$. The slope of the fitted line in this approach is the reddening $E(B-V)$, while the intercept is the extinction-free modulus $(m-M)_0$, which can be translated into the distance $d_0$. In the isochrone fitting process, where the distance is used as one of the constraints, the reproduced $d_0$ value is the one that is being compared with $d$.

\section{Radial Velocities}
The RVs for BD44,  extracted using TODCOR method, are given in Table.\ref{tab:RVBD44}.
\begin{table*}
 \caption{RV measurements of BD44.}
 \label{tab:RVBD44}
 \small
\begin{tabular}{ccccccccc}
\hline
BJD-2450000   & $v_1 \, (km\, s^{-1})  $    & $\epsilon_{1}\, (km\, s^{-1})  $ & $v_2 \,(km\, s^{-1})  $    & $\epsilon_2 \, (km\, s^{-1})  $ & $\gamma\, (km\, s^{-1})  $ & $\epsilon_{\gamma}\,(km\, s^{-1})  $ & $v_3\, (km\, s^{-1})  $    & $\epsilon_3 \,(km\, s^{-1}) $ \\
\hline
7022.350624 &  -21.499 & 0.697 & -146.291 & 0.505 &  -81.664 & 0.754 &  -78.406 & 0.079 \\
7061.170892 &   13.564 & 0.428 & -149.393 & 0.973 &  -65.001 & 0.956 & -106.205 & 0.100 \\
7062.263807 & ---      & ---   &   -3.669 & 0.863 & ---      & ---   & ---      & --- \\
7110.041913 &  -22.493 & 0.471 & -139.493 & 0.394 &  -78.901 & 0.654 &  -78.143 & 0.075 \\
7111.161334 & -160.274 & 0.578 &    3.675 & 0.355 &  -81.230 & 0.878 &  -76.942 & 0.086 \\
7142.999934 & -148.156 & 0.714 &  -23.570 & 0.426 &  -88.090 & 0.746 &  -63.215 & 0.083 \\
7144.008005 &  -13.377 & 0.565 & -161.962 & 1.019 &  -85.013 & 0.929 &  -62.731 & 0.135 \\
7148.097089 &  -16.490 & 1.381 & -160.184 & 0.379 &  -85.768 & 1.023 &  -62.282 & 0.077 \\
7490.046763 & -164.541 & 1.803 &   -2.964 & 0.496 &  -86.641 & 1.250 &  -68.064 & 0.227 \\
7526.015228 &  -10.156 & 2.387 & -159.806 & 0.391 &  -82.306 & 1.452 &  -76.503 & 0.095 \\
7528.026002 & -164.479 & 1.040 &    7.485 & 0.497 &  -81.571 & 1.032 &  -77.027 & 0.140 \\
7530.053976 &  -13.474 & 1.193 & -154.691 & 0.526 &  -81.558 & 0.965 &  -78.010 & 0.087 \\
7539.052136 & ---      & ---   &  -41.870 & 0.339 & ---      & ---   &  ---     & --- \\
7540.110845 &    3.205 & 0.606 & -166.138 & 0.514 &  -78.439 & 0.925 &  -82.127 & 0.123 \\
7755.261393 &   -6.755 & 0.688 & -164.869 & 0.429 &  -82.985 & 0.881 &  -69.878 & 0.103 \\
7813.057543 & -146.164 & 2.782 &    2.688 & 0.778 &  -74.399 & 1.660 &  -92.769 & 0.105 \\
7813.175027 & -137.988 & 1.525 &   -8.703 & 0.446 &  -75.657 & 1.037 &  ---     & --- \\
7814.125822 &   -3.179 & 1.464 & -141.592 & 0.745 &  -69.911 & 1.081 &  -93.820 & 0.093 \\
7816.093416 & -150.858 & 2.080 &   12.322 & 0.828 &  -72.185 & 1.402 &  -95.491 & 0.128 \\
7816.322403 & -152.826 & 2.013 &   13.918 & 0.420 &  -72.435 & 1.343 &  -95.617 & 0.126 \\
7846.107080 &   16.267 & 1.456 & -123.088 & 0.526 &  -50.919 & 1.051 & -139.779 & 0.231 \\
7891.177386 &  -14.944 & 1.876 & -163.681 & 0.491 &  -86.653 & 1.240 &  -66.297 & 0.118 \\
7892.950768 & -152.526 & 0.821 &  -12.576 & 0.450 &  -85.053 & 0.839 &  -66.027 & 0.078 \\
7894.027357 & ---      & ---   & -154.775 & 0.303 & ---      & ---   & ---      & --- \\
7950.023399 &   -3.531 & 0.670 & -173.935 & 0.750 &  -85.687 & 0.978 &  -62.241 & 0.097 \\
7954.983416 & -164.296 & 0.700 &   -3.259 & 0.266 &  -86.656 & 0.882 &  -62.650 & 0.101 \\
8066.362164 & -156.728 & 2.378 &   14.806 & 0.152 &  -74.028 & 1.496 &  -91.838 & 0.117 \\
\hline\end{tabular}
\end{table*}

 \section{Spot Parameters}
 The \PHOEBE{} model for BD44 included one spot on the Aa star and two on the Ab star. While we considered the most stable spot (spot-2 on Ab) for our calculations of differential rotation, the other parameters are important while modelling the LC and are given in Table.\ref{tab:spotparams}. The spot parameters vary over the three \TESS{} sectors: Sector-16 (S16), Sector-22 (S22), and Sector-49 (S49).   
\begin{table}
\centering
 \caption{Spot parameters for spots on primary (Aa) and secondary (Ab) of BD44.}
 \label{tab:spotparams}
 \begin{tabular}{lccc}
  \hline
   Parameters &  S16 & S22 & S49\\
  \hline

   $T^\mathrm{spot}_\mathrm{Aa}$ &0.9535& 0.9362 &1.04848\\
  $T^\mathrm{spot1}_\mathrm{Ab}$ &0.9922& 0.9806 &0.9545\\
 $T^\mathrm{spot2}_\mathrm{Ab}$ &0.8988& 0.9296 &0.8499\\
  $r^\mathrm{spot}_\mathrm{Aa}$\hspace{0.1cm}   [deg] &29.449 &29.846&33.943\\
  $r^\mathrm{spot1}_\mathrm{Ab}$  [deg]&31.9971 & 30.446 &11.390\\
  $r^\mathrm{spot2}_\mathrm{Ab}$  [deg]&42.846 & 37.311 &39.611\\
   $c^\mathrm{spot}_\mathrm{Aa}$\hspace{0.1cm} [deg]&80.735 & 69.893 & 80.735\\
  $c^\mathrm{spot1}_\mathrm{Ab}$ [deg]&90.977 & 89.047 &90.977\\
  $c^\mathrm{spot2}_\mathrm{Ab}$ [deg]&22.000& 35.745 &21.913\\
   $l^\mathrm{spot}_\mathrm{Aa}$\hspace{0.1cm} [deg]&180.000& 194.891   &84.0973\\
  $l^\mathrm{spot1}_\mathrm{Ab}$ [deg]&162.488&  196.044   &262.305\\
   $l^\mathrm{spot2}_\mathrm{Ab}$ [deg]&0.000& 29.975  &180.233\\
  
  \hline
 \end{tabular}
 \end{table}

\section{BF Fitting Tables}
The BF fitting was done on multiple spectra of different epochs. Though the profile of every fit was similar, the flux fraction of each component varied slightly, which was necessary to consider while spectral disentangling. This variation was noticed for $v\mathrm{sin}(i)$ too but their variations were not significant to consider during spectral analysis. The complete tables for these parameters are available in the online version.


\bsp	
\label{lastpage}
\end{document}